\documentclass[review]{elsarticle}
\usepackage{amsmath}
\usepackage{amssymb}
\usepackage{color}
\usepackage{titlesec}
\usepackage{booktabs, threeparttable, stackengine}
\usepackage{tabu}
\usepackage{tabularx}
\usepackage{graphicx}    
\usepackage{subfig}   
\usepackage{float} 
\usepackage{placeins}  
\usepackage{multirow}
\usepackage{array}
\usepackage{gensymb}
\usepackage{colortbl}
\usepackage{enumitem}
\usepackage{xcolor}
\newcolumntype{L}[1]{>{\raggedright\arraybackslash}p{#1}}
\setstackEOL{\#}
\setstackgap{L}{12pt}
\usepackage[font=small]{caption}
\extrafloats{100}
\usepackage{makecell}
\usepackage{mathrsfs}  
\usepackage{lineno,hyperref}
\biboptions{sort&compress}
\usepackage{draftwatermark}

\journal{Aerospace Science and Technology}

\SetWatermarkText{Preprint - Submitted to [Aerospace Science and Technology]}
\SetWatermarkScale{0.8}
\SetWatermarkColor[gray]{0.8}
\SetWatermarkAngle{90} 
\SetWatermarkHorCenter{2cm}  
\SetWatermarkVerCenter{14cm} 

\begin{document}
\begin{frontmatter}
\title{Operational Feasibility Analysis of a Cryogenic Active
Intake Device for Atmosphere-Breathing Electric
Propulsion}

\author[inst1]{Geonwoong Moon}
\author[inst1]{Youngil Ko}
\author[inst2]{Minwoo Yi}
\author[inst1]{Eunji Jun\corref{cor}}
\affiliation[inst1]{organization={Korea Advanced Institute of Science and Technology},
            postcode ={34141}, 
            state={Daejeon},
            country={Republic of Korea}}
\affiliation[inst2]{organization={Agency for Defense Development},
            postcode ={34186}, 
            state={Daejeon},
            country={Republic of Korea}}            
\ead{eunji.jun@kaist.ac.kr}
\cortext[cor]{Corresponding author}

\begin{abstract}
\noindent Atmosphere-breathing electric propulsion (ABEP) systems are emerging for orbit maintenance in very-low-Earth orbit (VLEO) by capturing atmospheric propellant \textit{in situ} using an intake device. A previous study proposed the cryocondensation-regeneration active intake device (CRAID) to significantly enhance intake performance. This study investigates the operational feasibility of CRAID. A conceptual prototype model (CPM) is presented to verify its feasibility, and numerical analyses demonstrate the practical operational sequences, required cryocooler capacity, intake performance, and flight envelope. The numerical analyses employ the direct simulation Monte Carlo (DSMC) method with a phase change model and a 0D analytical model for RF ion thrusters. A significant improvement in intake performance is estimated based on the practical sequences, with compression performance at least 1000 times higher than that of prevalent intake devices. The capability for consistent propellant supply is observed regardless of atmospheric conditions. A model satellite incorporating CPM confirms that CRAID enables complete drag compensation at altitudes above 190 km without limiting the upper boundary of the flight envelope.

\end{abstract}

\begin{keyword}
\texttt {Atmosphere-Breathing Electric Propulsion, Intake Device, DSMC, Very-Low-Earth-Orbit, Cryogenics}
\end{keyword}
\end{frontmatter}


\section{Introduction}
\begin{figure}[htb!]
	\centering
	\includegraphics[width=12cm]{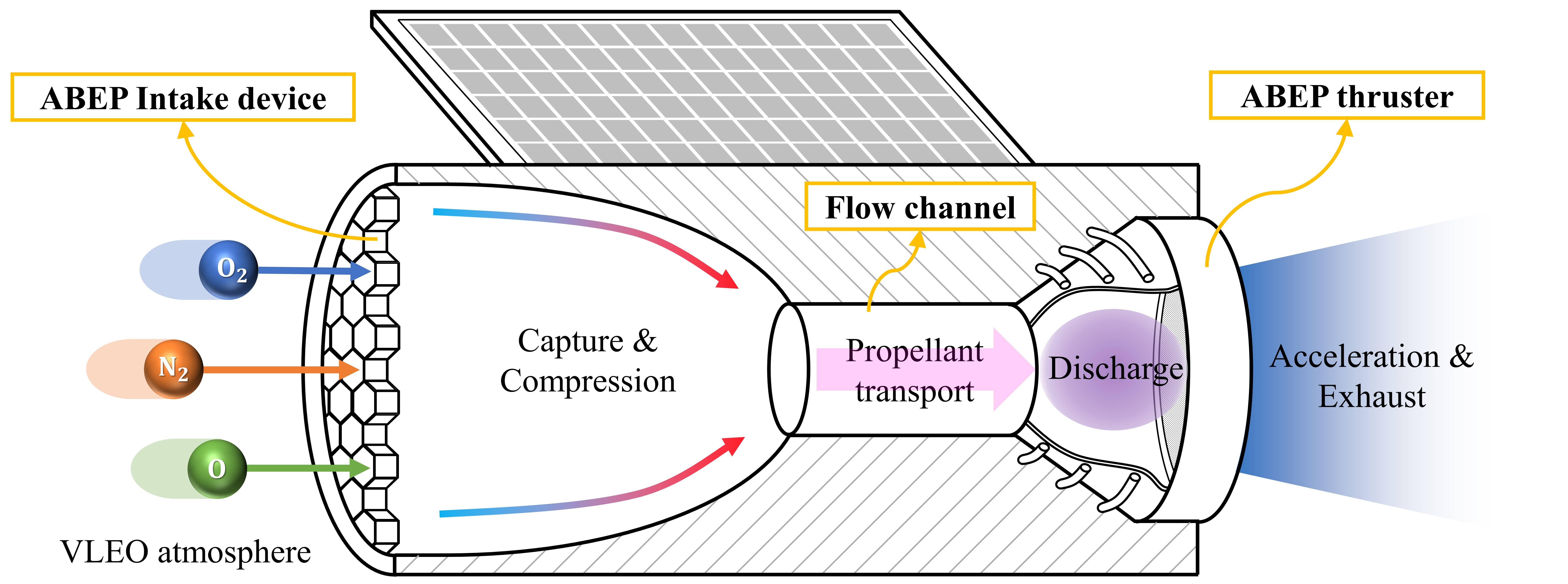}
	\caption{Schematic of ABEP working principles.}
	\label{fig:ABEP_concept}
\end{figure}

Very-low-Earth-orbit (VLEO) under 450 km offers numerous scientific and commercial benefits for satellite operation~\cite{Crisp2020}. However, the residual atmosphere exerts a significant drag force on spacecrafts and shortens their mission lifetimes~\cite{Schonherr2015}. Atmosphere-breathing electric propulsion (ABEP) is an emerging concept for efficient orbit maintenance in VLEO~\cite{Nishiyama2003,Filatyev2023,Moon2024ast}. Fig.~\ref{fig:ABEP_concept} illustrates the working principles of the ABEP system. Similar to air-breathing engines for aircraft, the ABEP system captures and compresses atmospheric flow in the ram direction using an intake device. The captured atmospheric gas is transferred to the ABEP thruster through a flow channel for thrust generation.\\ 


The ABEP systems incorporate an ABEP thruster and an intake device as two key components~\cite{Nishiyama2003,Filatyev2023,Andreussi2022}. First, ABEP thrusters have been considered based on prevalent EP mechanisms~\cite{Filatyev2023,Andreussi2022}. The radio-frequency ion thruster (RIT) is a promising candidate for ABEP due to its high specific impulse and electrodeless discharge. The stable operation of the RIT-based ABEP thruster has been demonstrated using $N_2$ and $O_2$ gas mixture~\cite{Cifali2011,Tisaev2022,Zorn2024}. Its discharge characteristics and thrust performance have been well described using 0D analytical models~\cite{Mrozek2021,Lopez-Uricoechea2022,Moon2023psst,Ko2023}. Second, the ABEP intake devices are designed to serve two functions: capture and compression. The intake device must effectively guide the VLEO atmospheric flow into the thrusters to ensure propellant capture, as the freestream flows at a relative orbital velocity of approximately 8 km/s with respect to the spacecraft. Moreover, pressurizing the captured gas is crucial for propellant processing in ABEP thrusters, as the VLEO atmosphere is highly rarefied. The performance of the ABEP intake device can be measured using the capture efficiency $\eta_c$ and the compression ratio $CR$~\cite{Moon2023vac}:
\begin{equation}
	\eta_c = \frac{\dot{N}_{out}}{\dot{N}_{in}} = \frac{n_{out}u_{out}A_{out}}{n_{\infty}u_{\infty}A_{in}}\label{eq:eta_c},
\end{equation}
\begin{equation}
	CR = \frac{n_{out}}{n_{\infty}}\label{eq:CR},
\end{equation}
where $\dot{N}$ represents the particle flow rate, and $n$ denotes the number density. The stream velocity is given by $u$, and $A$ refers to the cross-sectional area of the intake device. The subscript $\infty$ corresponds to freestream conditions, while $in$ and $out$ indicate the inlet and outlet of the intake device, respectively. The capture efficiency $\eta_c$ represents the flow rate ratio of the captured gas entering the thruster to the total atmospheric freestream incident on the intake device's inlet area. The compression ratio $CR$ represents the increase in the number density of the captured propellant relative to the freestream atmospheric number density.\\ 

Various designs of the intake device have been investigated to achieve high $\eta_c$ and $CR$~\cite{wu2022progress,Shoda2023,Nishiyama2003,Moon2023vac,Jackson2018,Romano2021,Zheng2021,Zheng2022,Rapisarda2023,Levchenko2020,Jin2024,Fontanarosa2024,Yakunchikov2025,Li2015,Moon2024ast,Yi2025,Moon2022}. Intake devices that incorporate a simple flow channel are classified as passive intake devices, which have been the focus of most previous studies~\cite{wu2022progress,Shoda2023,Nishiyama2003,Moon2023vac,Jackson2018,Romano2021,Zheng2021,Zheng2022,Rapisarda2023,Levchenko2020,Jin2024,Fontanarosa2024,Yakunchikov2025}. Previous studies have estimated that passive intake devices can achieve a wide range of $\eta_c$ from 10 to 95\%, depending on various shape and the gas-surface scattering characteristics of the surface material, while $CR$ ranges from 6 to 650. Due to their simple and rigid structure, passive intake devices can offer advantages in safe-life design and vibration resistance during space launch. However, several challenges remain in developing a feasible intake device that can be integrated into a complete ABEP system. First, improvement of $CR$ may be required to satisfy the discharge conditions of the thrusters. The average number density in VLEO decreases from approximately $10^{16}$ at 180 km to $10^{15}$ m$^{-3}$ at 300 km under normal solar activity~\cite{Ko2023,Emmert2020}. Both modeling and experimental studies have demonstrated that the conventional EP thrusters operate at a neutral gas number density of approximately $10^{19}$--$10^{20}$ m$^{-3}$ in the vicinity of the gas injection region~\cite{NAKAYAMA2014,Tsukizaki2021,Boeuf2017,Chabert2012,Grondein2016,Faraji2023}. Therefore, a $CR$ of 1000 or higher is preferable for sustaining discharge in ABEP thrusters. Second, incorporating a flow regulation system may be necessary to stabilize the propellant supply, as the VLEO atmospheric density and temperature fluctuate severalfold depending on local time, geographic location, and the solar activity phase at a constant altitude~\cite{Emmert2020}. Lastly, the large outlet diameter poses a challenge for integrating with conventional EP thrusters, which typically employ injectors less than a few centimeters in diameter~\cite{Kim2009,Lenguito2019,Snyder2009,Dietz2019,Langendorf2013}. Reducing the outlet-to-inlet diameter ratio could maximize the captured flow rate and improve thruster compatibility, given that passive intake devices currently have a ratio exceeding 20\%~\cite{wu2022progress,Moon2024ast}.\\

As an alternative solution, active intake devices have been proposed that incorporate driving mechanisms in addition to the passive intake devices. Li \textit{et al.} proposed an active intake device that adopts multi-hole plate and turbomolecular pump (TMP) mechanisms~\cite{Li2015}. Numerical simulations and subsystem experiments estimated a promising intake performance for the TMP-based active intake device, with an $\eta_c$ of approximately 60\% and a $CR$ of 3500 or higher~\cite{Li2015}. Moon \textit{et al.} proposed a cryogenic active intake device concept, called the cryocondensation-regeneration active intake device (CRAID)~\cite{Moon2024ast}. CRAID employs cryopumping mechanisms through repeated operational sequences to condense and regenerate atmospheric gas in VLEO. Numerical simulations predicted that CRAID could achieve an $\eta_c$ of 42\% with a $CR$ of approximately $10^{5}$ throughout the operational cycle. CRAID presents a promising approach for enhancing intake performance, featuring a narrow flow channel with an outlet-to-inlet diameter ratio of 5\%. However, its reliance on additional heat transfer processes and sequential operations introduces several challenges. First, the practical durations of each sequence should be determined considering system requirements to ensure sustained and repeatable operation. Second, a comprehensive power budget analysis is essential, accounting for both cryogenic temperature control and thruster operation. Finally, investigating the feasibility of a complete drag compensation maneuver is necessary for a spacecraft that provides sufficient propellant flow rate and power balance to support a CRAID-based ABEP system.\\


This study aims to present the advanced concept of CRAID and investigate its operational feasibility. A conceptual prototype model is presented as an illustrative design for an operable CRAID system, and its feasibility is verified through numerical analyses. It evaluates system requirements, including practical sequence durations and cryocooler capacity, along with the corresponding intake performance and the flight envelope. First, the propellant transport characteristics and required cryocooler capacity are computed using the direct simulation Monte Carlo (DSMC) method with a phase change model. Second, the durations of each operational sequence are determined based on the computed cryocooler capacity and gas transport characteristics, and the corresponding intake performance is evaluated. Finally, a power budget is estimated to support both CRAID and the thrusters, leading to the derivation of a spacecraft geometry that meets the system requirements. The spacecraft's flight envelope is determined to achieve complete drag compensation. The 3D DSMC method is employed for drag calculations, while a 0D analytical model for the RIT is used to estimate thrust.\\ 

The remainder of this paper is organized into four sections. Section~\ref{sec:concept} introduces the design concept and operating mechanisms of CRAID. Section~\ref{sec:method} describes the physical models and numerical methodologies employed, including the DSMC method and a 0D analytical model for RIT. Section~\ref{sec:result} presents the conceptual prototype model of CRAID and analyzes its operational feasibility. Finally, Section~\ref{sec:conclusion} summarizes the study's conclusions and outlines directions for future research.\\


\section{Design Configuration and Operating Mechanisms of CRAID}
\label{sec:concept}

CRAID adopts the working principles of cryopumps commonly used in high-vacuum test facilities~\cite{Moon2024ast}. The cryopumps condense and trap gas molecules on cryopanels cooled to 10–20 K during the vacuum test, a process known as the condensation sequence~\cite{Day2007}. At 20 K, most atmospheric constituents, including $N_2$ and $O_2$, exhibit extremely low saturated vapor pressures of less than $10^{-10}$ torr~\cite{brown1980vapor}. At the end of the vacuum test, the cryopump switches to the regeneration sequence, raising the temperature of the cryopanels and sublimating the condensed particles back into the gas phase.\\ 



\begin{figure}[htb!]
	\centering
	\subfloat[Conceptual image.]{\includegraphics[width=11.8cm]{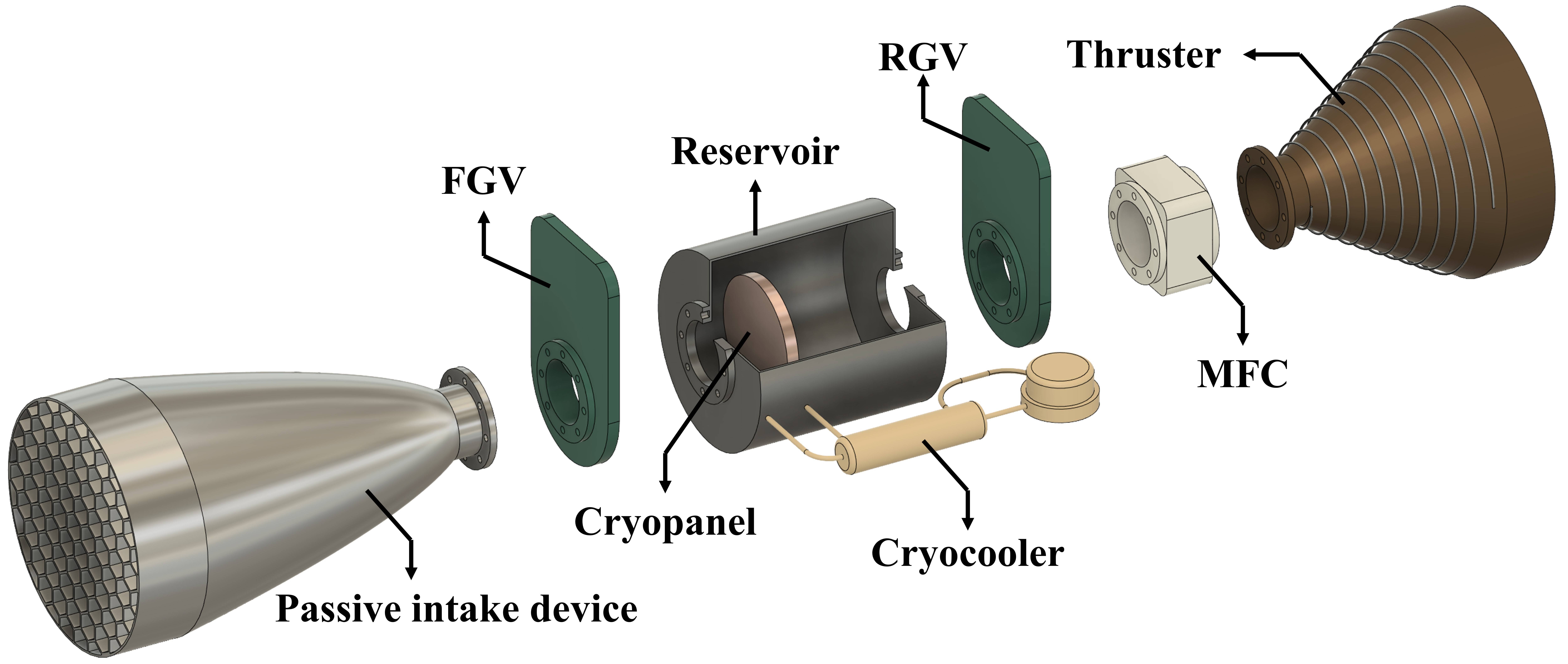}\label{fig:CRAID_concept}}
	\vfill
	\subfloat[Schematic diagram.]{\includegraphics[width=11.8cm]{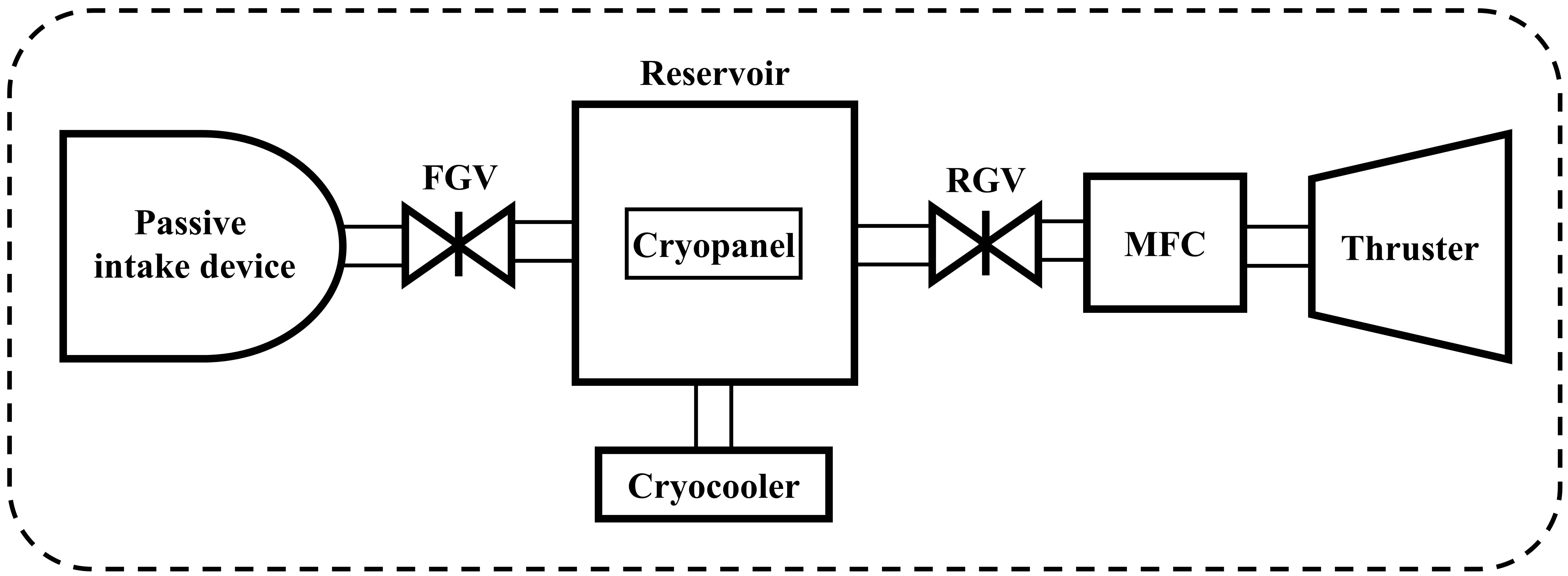}\label{fig:CRAID_schematic}}
	\caption{Design components of CRAID~\cite{Moon2024ast}.}
	\label{fig:CRAID_design}
\end{figure}

\begin{figure}[h!] 
	\centering
	\subfloat[Operational sequences]{\includegraphics[width=9.5cm]{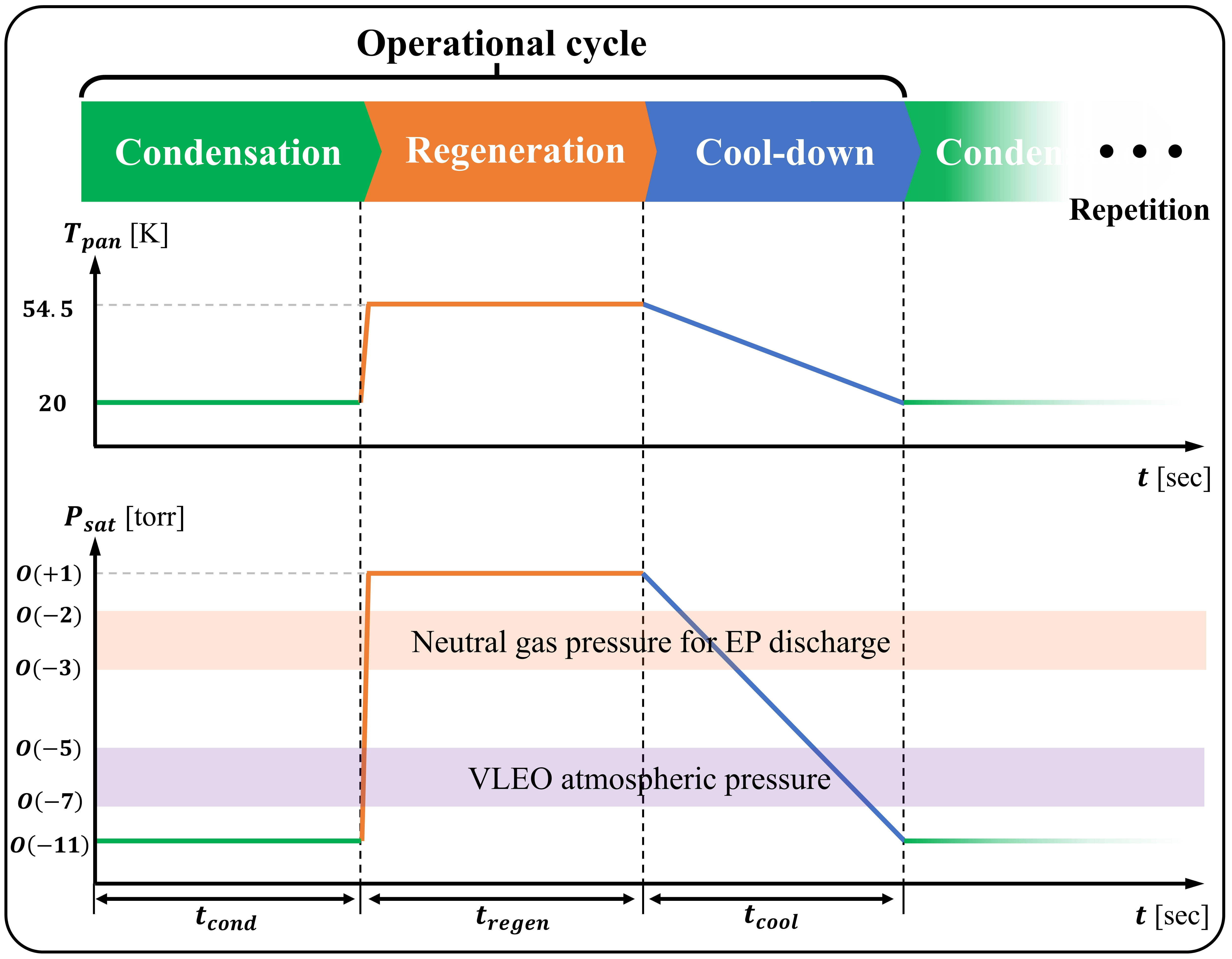}\label{fig:CRAID_sequence_schematics}}
	\vfill
	\subfloat[Working principles]{\includegraphics[width=9.5cm]{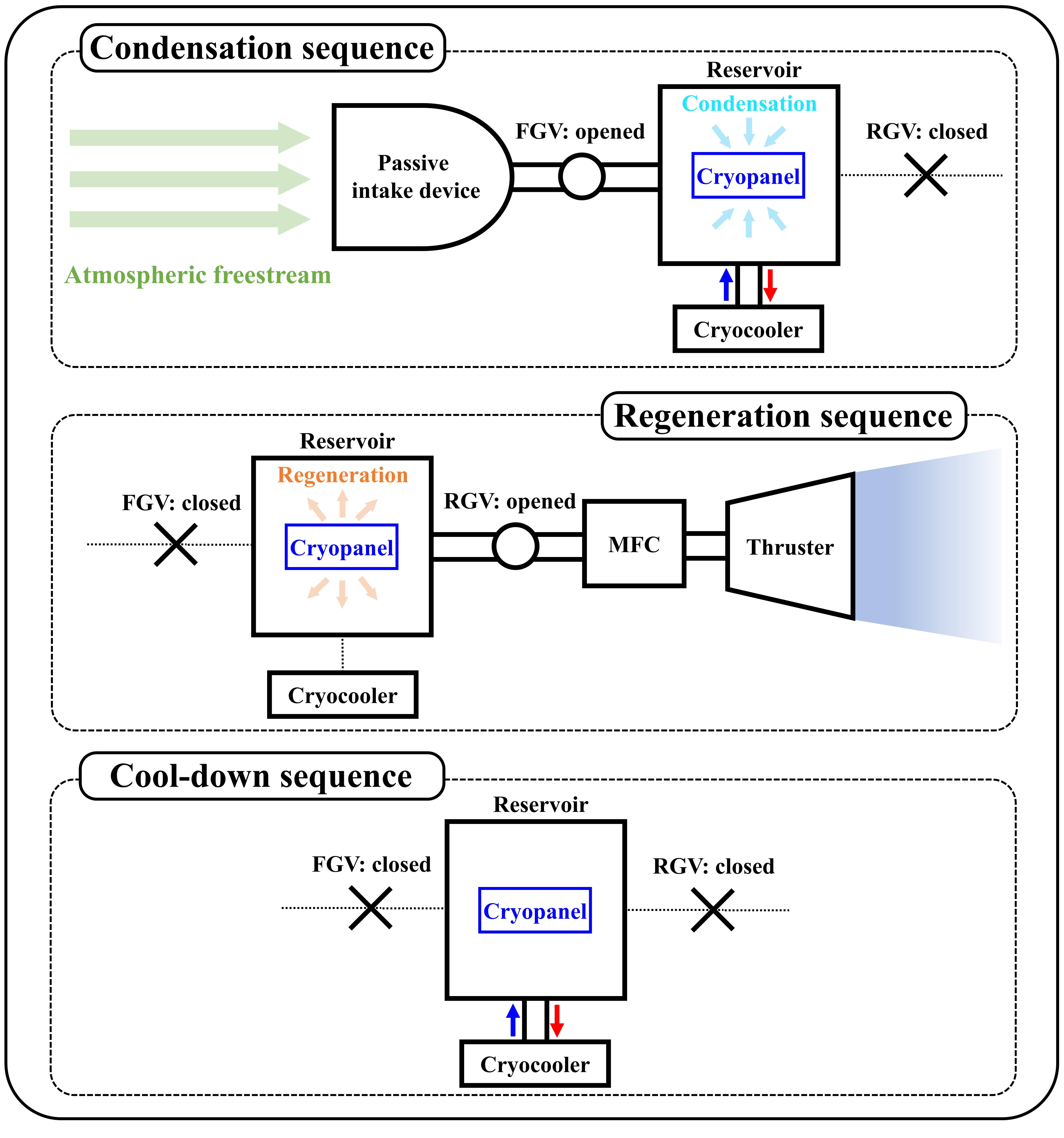}\label{fig:CRAID_working}}
	\caption{Operational sequences and working principles of CRAID.}
	\label{fig:CRAID_sequence}
\end{figure}

CRAID is designed with a temperature-adjustable cryopanel for condensing gas molecules, along with a conventional passive intake device~\cite{Moon2024ast}. The design components are depicted in a conceptual exploded view in Fig.~\ref{fig:CRAID_design}(a), and the schematic diagram of the assembly is shown in Fig.~\ref{fig:CRAID_design}(b).
CRAID consists of five main components; passive intake device, cryocooler, cryopanel, reservoir, and two gate valves. A mass flow controller (MFC) and a thruster can be integrated into CRAID to form a complete ABEP system. The operational cycle of CRAID consists of three sequences: the condensation sequence, the regeneration sequence, and the cool-down sequence. The operational sequences with a targeted cryopanel temperature $T_{pan}$ and the corresponding saturated vapor pressure $P_{sat}$ are shown in Fig.~\ref{fig:CRAID_sequence}(a). The functioning of each design component in achieving these sequences is illustrated in Fig.~\ref{fig:CRAID_sequence}(b).
During the condensation sequence lasting for a duration of $t_{cond}$, the frontal gate valve (FGV) is open to direct the atmospheric freestream into the reservoir. Within the reservoir, the cryopanel temperature is maintained at 20 K using a cryocooler. At this temperature, the saturated vapor pressure of atmospheric constituents is significantly lower than the atmospheric pressure in VLEO, allowing the captured gas molecules to condense onto the cryopanel. Once the desired amount of gas has condensed, the regeneration sequence proceeds for a duration of $t_{regen}$. During the regeneration sequence, the cryopanel temperature is increased to 54.5 K, staying below the triple point of $N_2$ and $O_2$. At this elevated temperature, the condensed molecules sublimate and are regenerated as gaseous propellant within the reservoir. The pressure of the regenerated gas can be increased up to a saturated vapor pressure of a few torr, which exceeds the neutral gas pressure typically required for EP discharge. During the regeneration sequence, FGV is closed, and the rear gate valve (RGV) is opened to transfer the regenerated gas to the thruster for thrust generation. Once the regenerated gas is fully depleted during thruster operation, the cool-down sequence begins. During the cool-down sequence, both the FGV and RGV are closed while the cryopanel is cooled to 20 K for a duration of $t_{cool}$, thereby preparing the system for the next operational cycle.

\FloatBarrier


\section{Physical Models and Numerical Methodology}
\label{sec:method}

\subsection{Boltzmann equation}

The VLEO atmospheric freestream is a highly rarefied gas flow where the Knudsen number $Kn$ exceeds 10. This rarefied flow can be described by the Boltzmann equation (BE): 
\begin{equation}
	\frac{\partial}{\partial t}(nf) + \vec{c}\cdot \frac{\partial}{\partial \vec{r}}(nf)+\frac{\vec{F}}{m_p}\cdot\frac{\partial}{\partial \vec{c}}(nf)=\int_{-\infty}^{\infty}\int_{0}^{4\pi}n^2[f^*f^*_1-ff_1]c_r\sigma d\Omega d\vec{c}_1 .
	\label{eq:boltzmann}
\end{equation} 
The BE describes the evolution of the particle distribution function $f$ in phase space of position $\vec{r}$ and velocity $\vec{c}$. It incorporates the binary collision operator on the right-hand side, accounting for the distribution functions $f$ and $f_1$ of colliding particle pairs. In Eq.~(\ref{eq:boltzmann}), $t$ denotes time, $\vec{F}$ represents the external force, $m_p$ is the particle mass, $n$ is the particle number density, $c_r$ is the relative velocity, $\sigma$ is the collision cross-section, $\Omega$ is the solid angle, and the asterisk represents the post-collisional distribution function. 

\subsection{DSMC method}

The direct simulation Monte Carlo (DSMC) method is a particle-based simulation technique that provides a stochastic solution to the BE~\cite{bird1994molecular}. The DSMC method calculates the time evolution of simulation particles, while accounting for intermolecular collisions and gas-surface interactions (GSI). The variable soft-sphere (VSS) model is used to calculate the collision cross-section $\sigma$ for intermolecular collisions~\cite{bird1994molecular,Moon2024ast}. The GSI is modeled using the Maxwell model, which describes the kinetic energy of reflected particles through the energy accommodation coefficient $\alpha_\epsilon$: 
\begin{equation} 
    \alpha_\epsilon = \frac{T_{k,pre} - T_{k,post}}{{T_{k,pre}}-T_{s}}, 
    \label{eq:alpha} 
\end{equation} 
where $T_{k,pre}$ and $T_{k,post}$ represent the kinetic energy of a particle before and after the GSI, respectively, while $T_{s}$ denotes the surface temperature. A specular reflection occurs when $\alpha_\epsilon = 0$, whereas a fully diffuse reflection takes place when $\alpha_\epsilon = 1$.

\subsection{Phase change model}

The phase change model for sublimation and condensation is incorporated into the DSMC method. Sublimation is described using the Hertz-Knudsen (HK) equation~\cite{Kolasinski,Moon2024ast}: 
\begin{equation} 
    J=\frac{p}{\sqrt{2\pi m_p k_B T_{s}}}, 
    \label{eq:HK} 
\end{equation} 
where $J$ is the sublimating particle flux, $p$ is the gas pressure, and $k_B$ is the Boltzmann constant. The $J$ at a given $T_s$ is determined by substituting $p$ with the empirical formula for the saturation pressure $p_{sat}$~\cite{brown1980vapor,Moon2024ast}: 
\begin{equation} 
    \ln p_{sat} = A_0 + \frac{A_1}{T_{s}} + \frac{A_2}{T_{s}^2} + \frac{A_3}{T_{s}^3} + \frac{A_4}{T_{s}^4}. 
    \label{eq:sat} 
\end{equation} 
Here, $p_{sat}$ is given in units of torr, and the coefficients $A_0$ through $A_4$ are empirically determined~\cite{brown1980vapor,Moon2024ast}. Condensation is characterized by the sticking coefficient $\alpha_s$, which quantifies the probability of an incident gas particle condensing upon a surface element. For the cryopanel surface, $\alpha_s$ is assumed to be unity~\cite{Welch2001,Bisschop2006,Eisenstadt1970,Dawson1965}. The DSMC code with the phase change model is adopted from Moon \textit{et al.}, whose code was developed based on the SPARTA open-source DSMC framework~\cite{Moon2024ast,Plimpton2019}.  

\subsection{0D analytical model for RIT}

The 0D analytical model for RIT calculates the exhaust ion beam current and corresponding thrust based on input parameters of thruster specifications, propellant flow rate, and total input power. The model solves a set of simultaneous equations that govern the continuity, energy balance, ambipolarity, and charge quasi-neutrality of plasma constituents. This study adopts the 0D analytical model code developed by Ko \textit{et al}.~\cite{Ko2023}, which can calculate RIT performance using an $N_2$-$O_2$ mixture propellant.

\clearpage

\section{Result and Discussion}
\label{sec:result}

\subsection{Framework for the operational feasibility analysis of CRAID}


The feasible operation of CRAID requires identifying the system requirements, including practical sequence durations and cryocooler specifications, along with an appropriate operable altitude range. The practical sequence durations $t_{cond}^*$, $t_{regen}^*$, and $t_{cool}^*$ are necessary to maintain a repeatable operational cycle. Determining the appropriate specifications of the cryocooler, including cooling capacity $Q_{cool}$ and input power $P_{cool}$, is crucial for executing the operational sequences. The flight envelope $h^*$ represents the altitude range for complete drag compensation, demonstrating the operational feasibility of CRAID in orbit.\\ 

\begin{figure}[htb!]
	\centering
	\includegraphics[width=\textwidth]{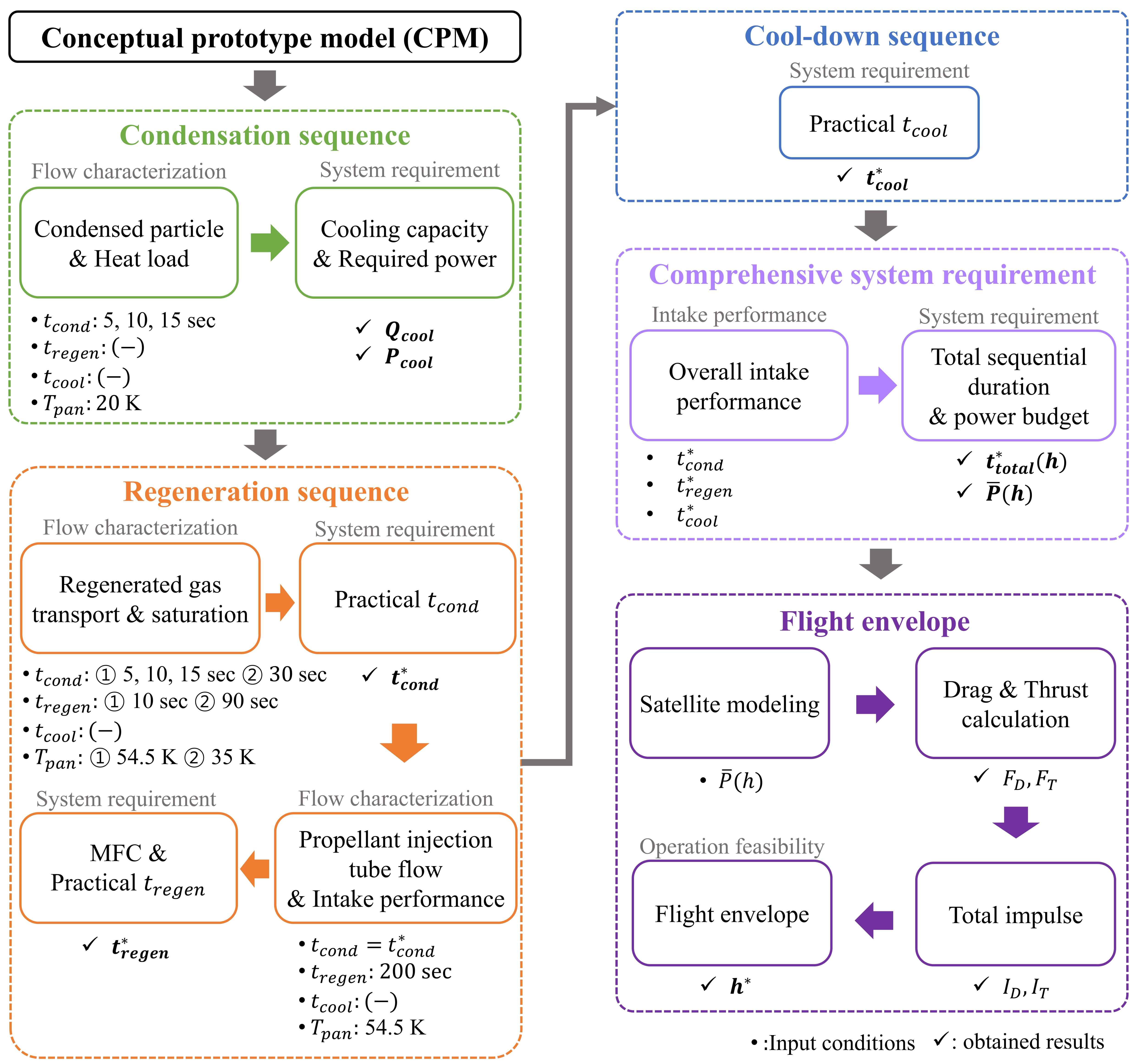}
	\caption{Framework for the operational feasibility analysis of CRAID.}
	\label{fig:strategy}
\end{figure}

A multi-step analytical framework is established to evaluate the operational feasibility of CRAID. The framework, depicted in Fig.~\ref{fig:strategy}, is structured around three operational sequences, comprehensive system requirement, and flight envelope. As a prerequisite, a conceptual prototype model (CPM) of CRAID is developed as an object of the operational feasibility analysis detailed below:

\begin{enumerate}
\item Condensation sequence
    \begin{itemize}
        \item The gas flow during the condensation sequence is characterized using DSMC simulation over an arbitrary $t_{cond}$ range of 5 to 15 seconds.
        \item The heat load on the cryopanel is calculated, and the required $Q_{cool}$ and $P_{cool}$ are determined. 
    \end{itemize}

\item Regeneration sequence
    \begin{itemize}
        \item The transport of regenerated gas is simulated with $t_{regen} = 10$ seconds for each case of $t_{cond}=$5, 10, and 15 seconds. The saturated behavior of the regenerated gas is characterized using an additional case at an arbitrary $T_{pan}=35$ K, in which $t_{cond}$ is extended to 30 seconds and $t_{regen}$ to 90 seconds.
        \item Based on the transport and saturation characteristics of the regenerated gas, a practical $t_{cond}^*$ is determined to maximize intake performance.
        \item The flow through the propellant injection tube is characterized using condensed particle densities at $t_{cond}^*$, estimating the corresponding intake performance during $t_{regen}=200$ seconds.
        \item A practical $t_{regen}^*$ is determined by incorporating an MFC for flow regulation.   
    \end{itemize}

\item Cool-down sequence
    \begin{itemize}
        \item A practical $t_{cool}^*$ is calculated based on $Q_{cool}$ and the heat capacity of the cryopanel.
    \end{itemize}

\item Comprehensive system requirement
    \begin{itemize}
        \item By aggregating the practical sequence durations to obtain $t_{total}^*$, the overall intake performance and the required total power budget $\bar{P}$ for CPM are determined.
        \item The variations in $t_{total}^*$ and $\bar{P}$ at different altitudes $h$ are investigated.
    \end{itemize}
    
\item Flight envelope
\begin{itemize}
    \item A satellite geometry is modeled to satisfy the system requirements at altitudes between 150 and 300 km.
    \item The drag and thrust acting on the satellite are calculated using the DSMC method and a 0D analytical model for RIT, respectively.
    \item The total impulses of thrust and drag during the operational cycle are compared.
    \item The operable range of $h^*$ for complete drag compensation is determined.
\end{itemize}
\end{enumerate}

\subsection{Conceptual prototype model of CRAID}

\begin{figure}[h!]
	\centering
	\includegraphics[width=12cm]{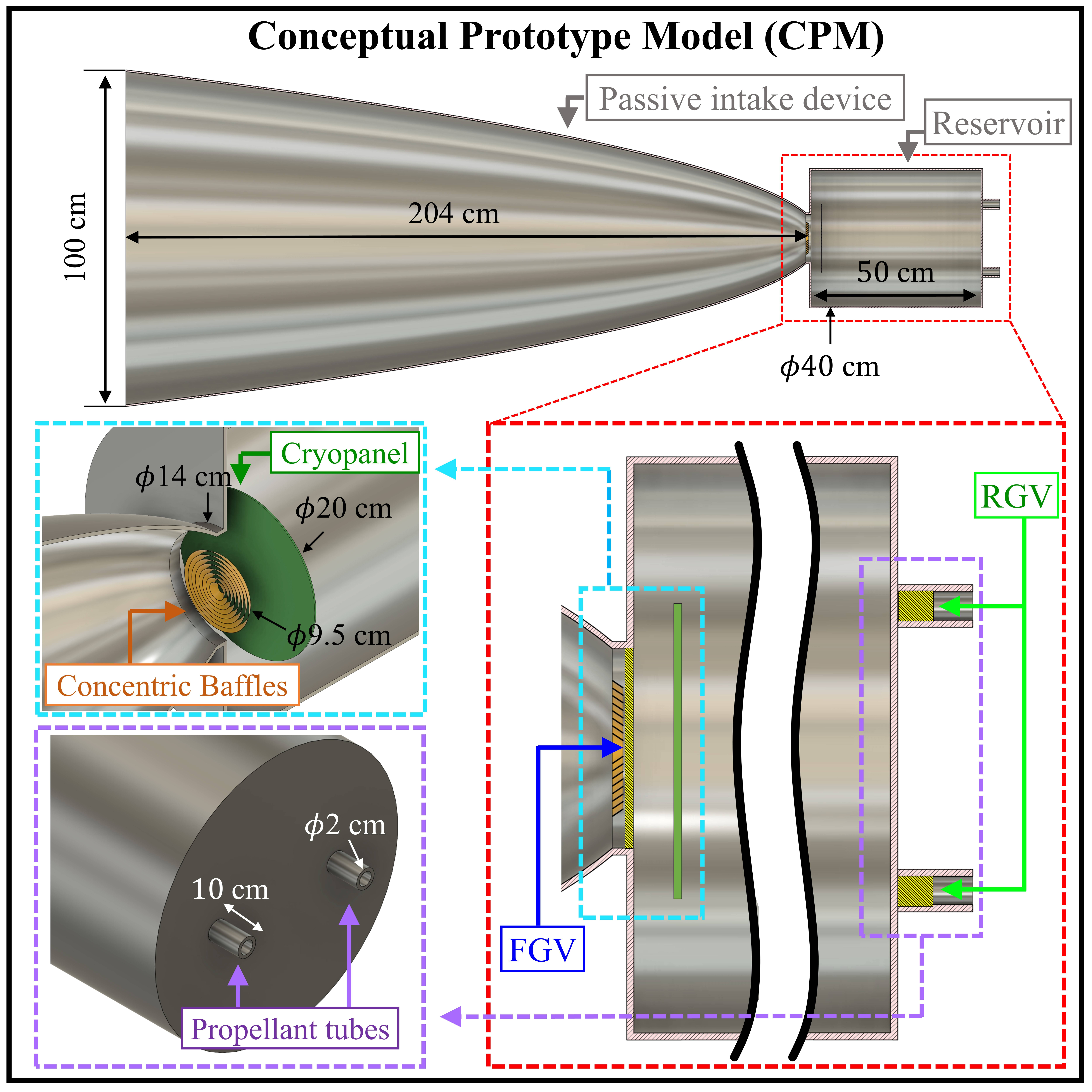}
	\caption{Conceptual prototype model (CPM) of CRAID.}
	\label{fig:CRAID_simple}
\end{figure}

A conceptual prototype model of CRAID, referred to as CPM, is designed for the operational feasibility analysis. Fig.~\ref{fig:CRAID_simple} illustrates the design configuration of CPM. A parabolic intake device with a diameter of 1 m is employed to enhance capture efficiency through the focusing effect~\cite{Moon2023vac,Zheng2022}. The outlet of the passive intake device is connected to a reservoir through an aperture with a diameter of 14 cm. At the interface between the passive intake device and the reservoir, the FGV and concentric baffles are positioned. Direct impingement of high-speed freestream particles on the cryopanel may lead to a high heat load and the detachment of condensed particles. Concentric baffles are used to enhance the thermalization of the captured gas and mitigate the direct impingement of freestream particles, while reducing external radiative heat transfer. The reservoir has a diameter of 40 cm and a length of 50 cm, providing an inner volume $V_{res}$ of 0.12 m$^{3}$. A disk-shaped cryopanel with a diameter of 20 cm and a thickness of 2 mm is installed inside the reservoir. The thin disk-shaped design can reduce the heat capacity of the cryopanel, allowing for a shorter cool-down sequence duration. Its wide frontal area covers the entire reservoir entrance in the ram direction, ensuring a high condensation probability for the incoming flow during the condensation sequence. Two propellant injection tubes are connected at the end of the reservoir to supply the regenerated propellant into twin RITs. Each propellant injection tube has a diameter of 2 cm and is equipped with an RGV. Correspondingly, the outlet-to-inlet diameter ratio is 2\% in CPM, which is less than 1/10 that of the conventional passive intake devices~\cite{wu2022progress,Moon2024ast}.

\subsection{Condensation sequence}

\subsubsection{Gas transport characteristics}
\label{subsec:cond1}
\begin{figure}[h!] 
	\centering
	\subfloat[Condensation sequence.]{\includegraphics[width=8cm]{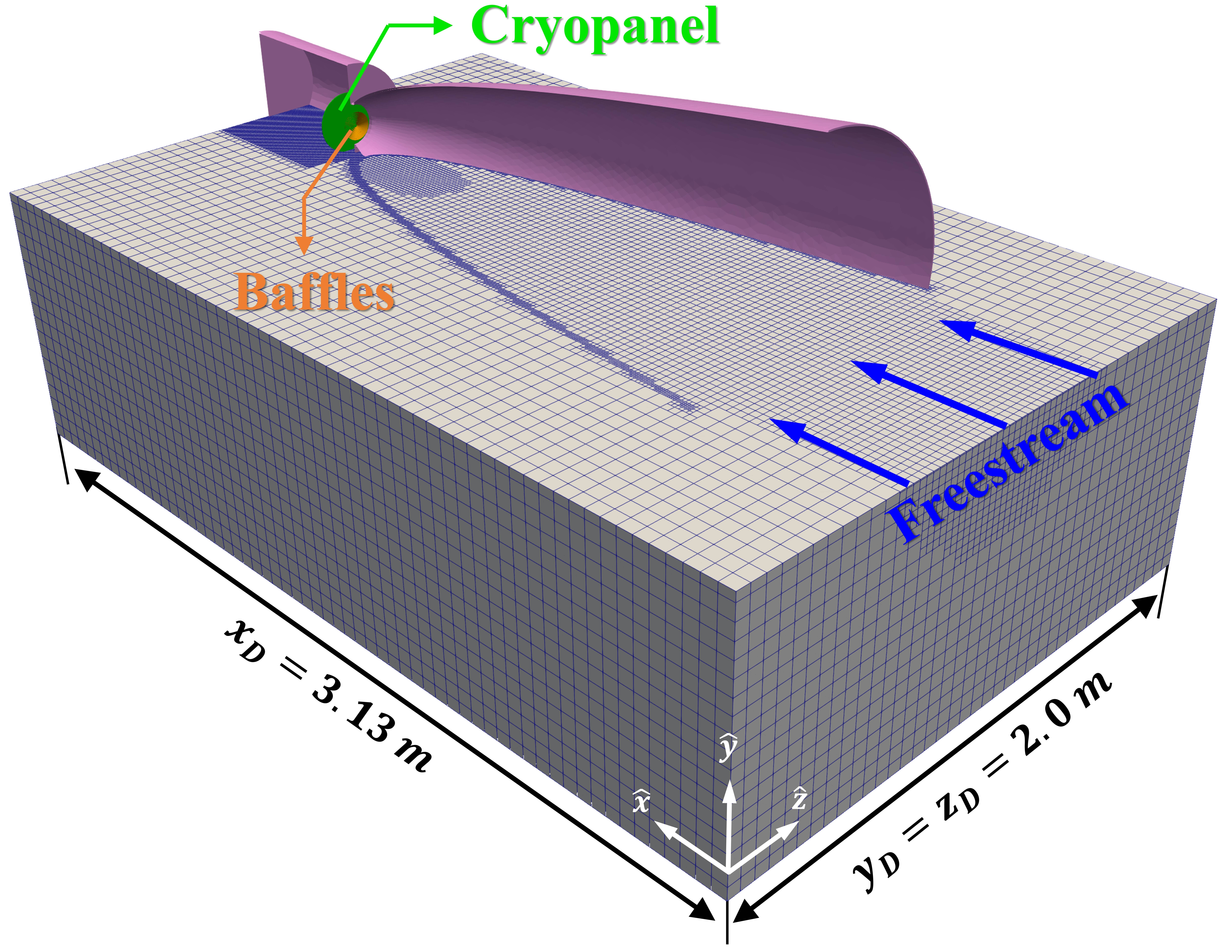}\label{fig:domain_cond}}
	\vfill
	\subfloat[Regeneration sequence.]{\includegraphics[width=8cm]{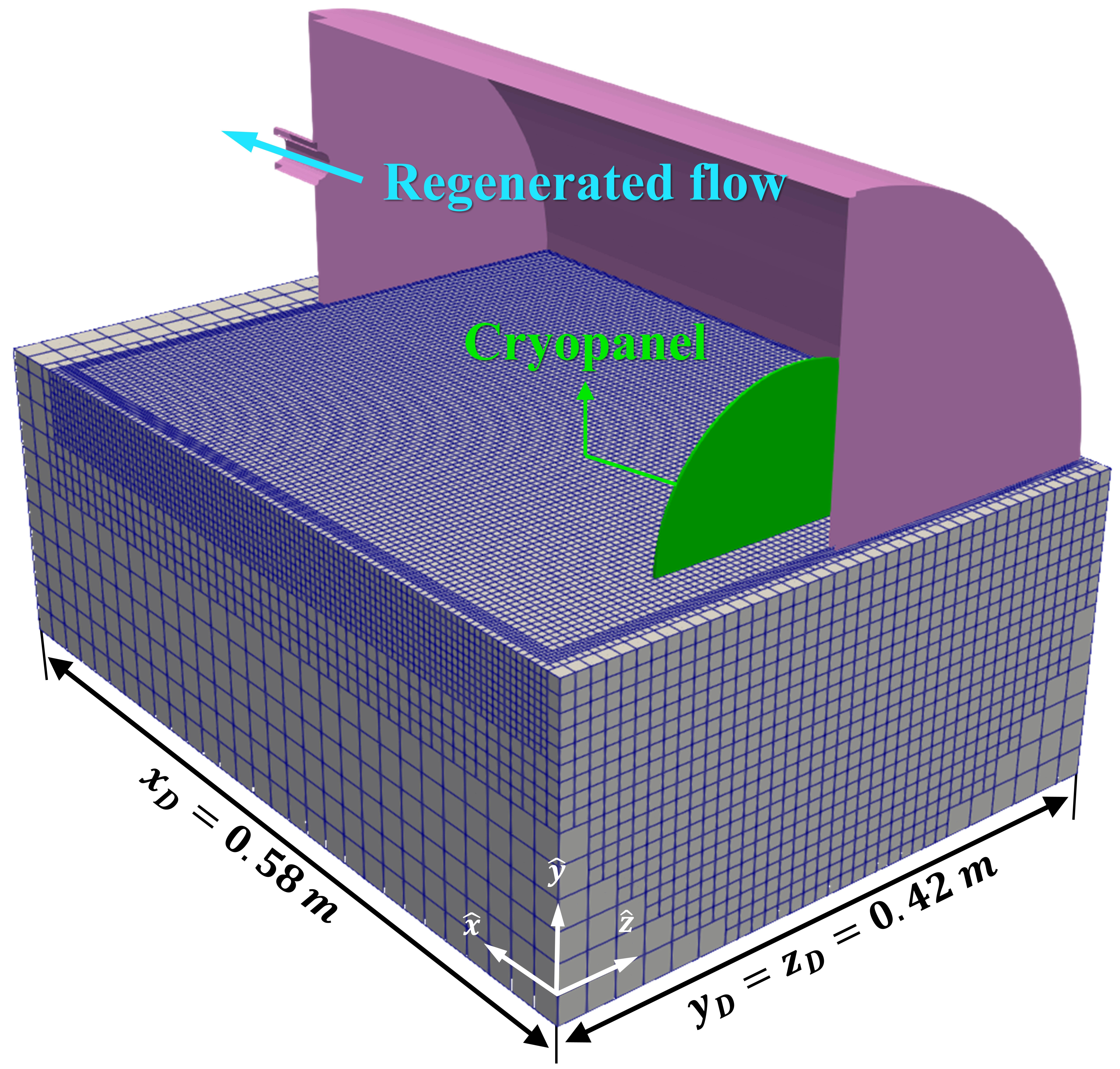}\label{fig:domain_regen}}
	\caption{Simulation domain and surface configuration according to operating sequences.}
	\label{fig:domain}
\end{figure}

The gas flow during the condensation sequence is simulated using the DSMC method incorporating a phase change model. The calculations are performed in the 3D domain shown in Fig.\ref{fig:domain_cond}. The simulation domain is divided into cubic calculation grid cells using the 5-level octree adaptive mesh refinement (AMR) technique. CPM illustrated in Fig.~\ref{fig:CRAID_simple} is implemented in the simulation domain.
The propellant injection tubes are excluded from the calculation, as the RGV blocks gas flow during the condensation sequence. A fully specular GSI model is applied to the passive intake device to maximize capture efficiency~\cite{wu2022progress,Zheng2022,Moon2023vac}. The baffles, reservoir, and cryopanel are assigned a diffuse GSI model with a surface temperature of 54.5 K. Additionally, a phase change model is applied to the cryopanel at 20 K. The boundaries of the computational domain are set as open, and atmospheric particles are generated at these boundaries according to the prescribed freestream conditions. The freestream conditions correspond to the orbital environment at an altitude of 200 km and are obtained using the NRLMSIS 2.0 model, assuming normal solar activity and a global average over equidistant points along latitude and longitude~\cite{Emmert2020}. At an altitude of 200 km, more than 99\% of the atmospheric constituents consist of $N_2$, $O_2$, and atomic oxygen (AO). Previous studies have reported that AO particles pressurized by the passive intake device can significantly recombine into $O_2$ molecules and that AO adsorption-mediated surface chemistry can further enhance this recombination~\cite{Singh2014,Tagawa2013,Cartry2000,Ko2024,Moon2024ast}. Since the current study focuses on the operational feasibility analysis of CRAID, the freestream conditions are simplified by assuming complete recombination of AO atoms in the atmosphere. The $O_2$ number density listed in Table~\ref{tab:freestream} accounts for the recombined molecules, satisfying stoichiometric conditions. The condensation sequence simulation is conducted with a time step size of $5\times10^{-6}$ seconds and a total of 3 million steps, corresponding to $t_{cond}$ of 15 seconds. The macroscopic properties of the flow field are tallied for every 0.1 million steps. The condensed particle density on each cryopanel's surface element is recorded every 1 million steps, corresponding to 5 seconds of physical time.\\

\begin{table} [!htb]
	\begin{center}
		\caption{Freestream conditions.}
		\begin{tabular}{c c c c }
			\toprule
			$\mathrm{Species}$ & $n_\infty$ [$\mathrm{m^{-3}}$] & $T_\infty$ [K] & $v_\infty$ [m/s]  \\ \midrule
			$O_2$    & $1.72\times10^{15}$       & \multirow{2}{*}{830}       & \multirow{2}{*}{7784}	 \\
			$N_2$    & $3.04\times10^{15}$       &     &  	\\
			\bottomrule 
		\end{tabular}
		\label{tab:freestream} 
	\end{center}
\end{table}

\begin{figure}[htb!]
	\centering
	\includegraphics[width=8cm]{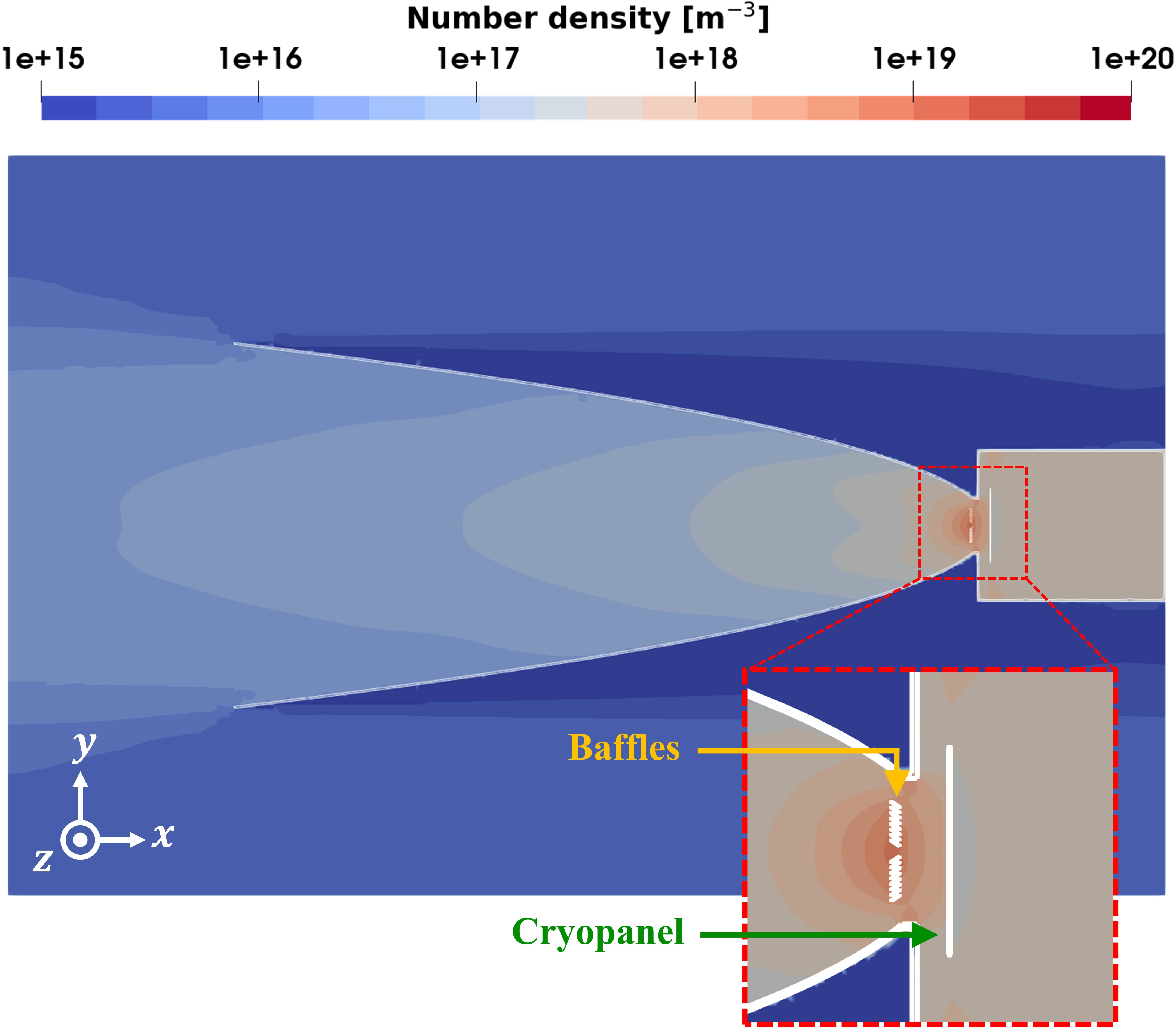}
	\caption{Number density contours during the condensation sequence.}
	\label{fig:cond_flow_contour}
\end{figure}

\begin{figure}[htb!]
	\centering
	\includegraphics[width=12cm]{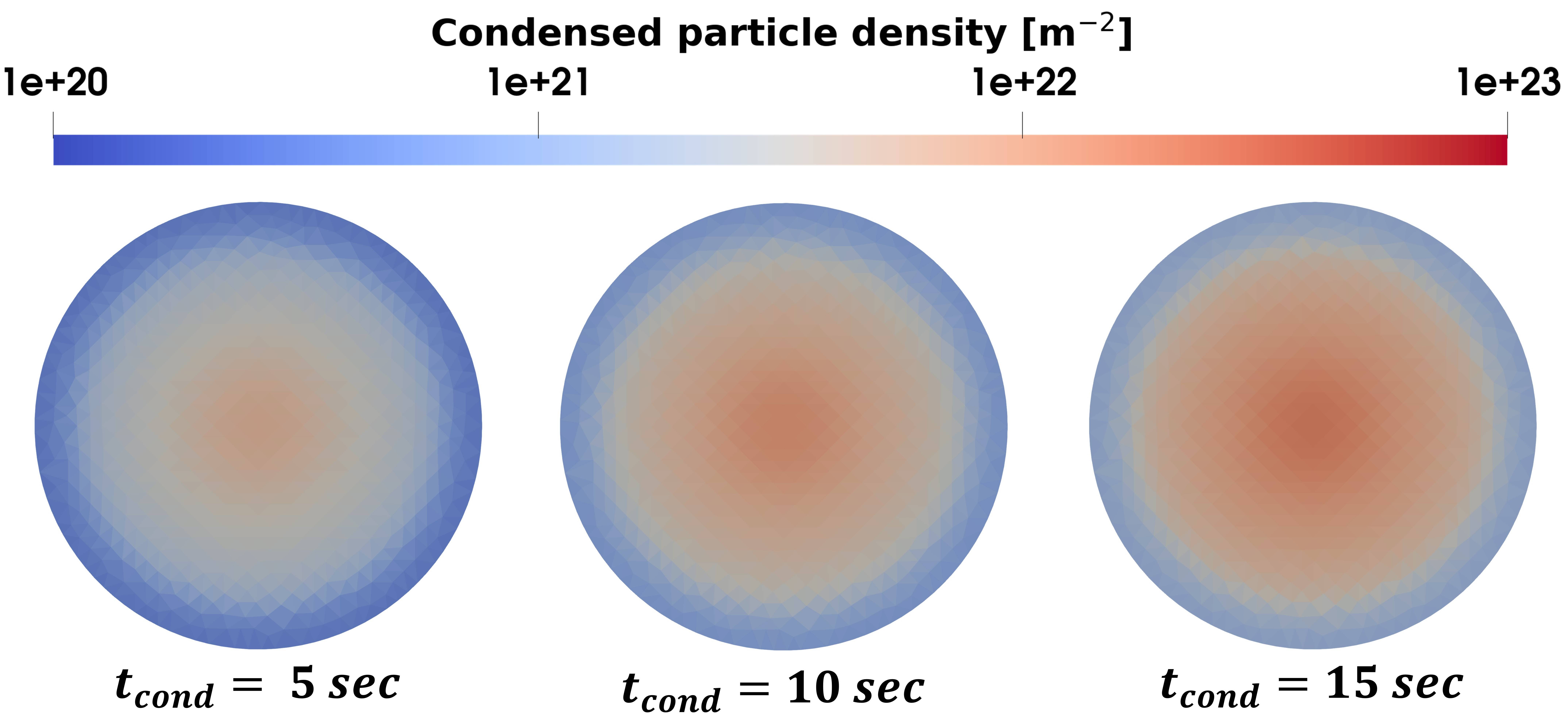}
	\caption{Distribution of condensed particle density $n_{s,c}$ on the cryopanel.}
	\label{fig:cond_surf_contour}
\end{figure}

\begin{figure}[htb!]
	\centering
	\includegraphics[width=8.5cm]{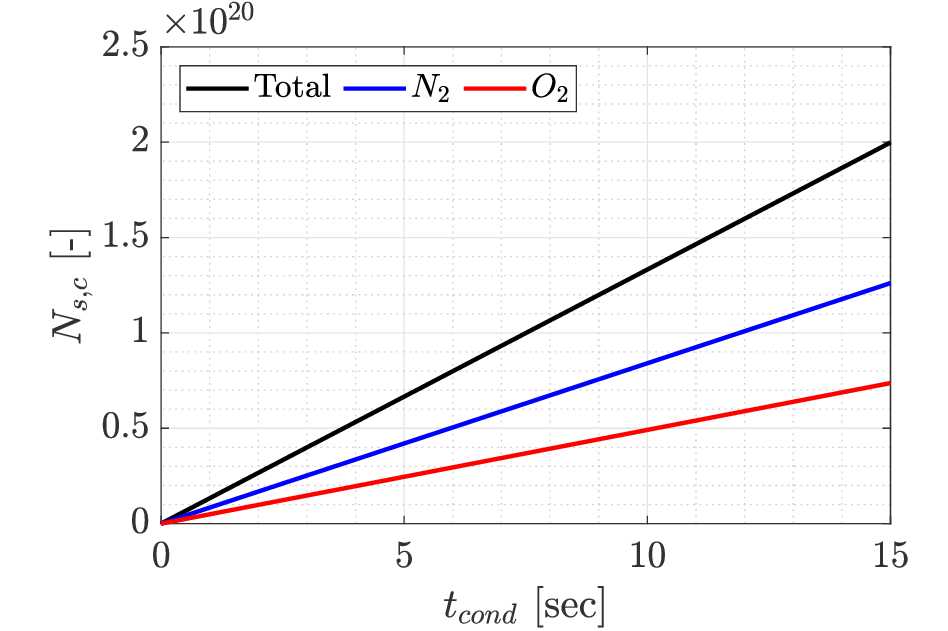}
	\caption{Number of condensed particles $N_{s,c}$ according to $t_{cond}$.}
	\label{fig:cond_total_part}
\end{figure}

During the condensation sequence, the flow field around CPM maintains a constant distribution regardless of the elapsed time. Fig.~\ref{fig:cond_flow_contour} shows the distribution of gas particle number density.
The passive intake device directs the flow toward a focal point where the baffles are positioned to thermalize the gas particles. The compressed gas molecules near the baffles are observed, with a decrease in density near the cryopanel, indicating condensation. Fig.~\ref{fig:cond_surf_contour} shows the distribution of the condensed particle density $n_{s,c}$ on the cryopanel, increasing with $t_{cond}$ at 5 second intervals.
Due to the axisymmetric and tapered geometry of the passive intake device, $n_{s,c}$ is highest at the center of the cryopanel. At $t_{cond} = 5$ seconds, $n_{s,c}$ peaks at $7.0 \times 10^{21}$ m$^{-2}$ at the center and decreases to $3.5 \times 10^{20}$ m$^{-2}$ near the edge. As the condensation sequence proceeds to $t_{cond}=15$ seconds, the local $n_{s,c}$ increases to $2.1 \times 10^{22}$ at the center and $1.1 \times 10^{21}$ near the edge. Due to the directional incidence of the gas, the particle flux impinging on the cryopanel's rear surface is lower than that on the front surface. Very low condensation occurs on the rear side, with $n_{s,c}$ remaining below $5 \times 10^{20}$ m$^{-2}$ for 15 seconds of $t_{cond}$. The total number of condensed particles $N_{s,c}$ at time $t$ is obtained by integrating $n_{s,c}$ over the cryopanel's surface area $A_{pan}$: 
\begin{equation}
    N_{s,c}(t) = \oint_{A_{pan}} n_{s,c}(t)\cdot dA,
\end{equation}
as shown in Fig.~\ref{fig:cond_total_part}. Both condensed $N_2$ and $O_2$ particles exhibit a linear increase with $t_{cond}$. The chemical composition of the condensed particles remains constant over time, with 63\% $N_2$ and 37\% $O_2$, aligning with the freestream composition. The rate of increase of total $N_{s,c}$ is $1.3 \times 10^{19}$ particles per second. The effective condensing-capture efficiency $\zeta_{cc}$ is introduced to assess the particle capture and storage capability of the passive intake device–baffles–cryopanel assembly. It is defined as the probability that a particle incident on the inlet area of CRAID will be condensed onto the cryopanel, expressed as:
\begin{equation}
    \zeta_{cc} = \frac{\dot{N}_{s,c}}{n_\infty u_\infty A_{in}}.
\end{equation}
Using the freestream conditions of 200 km VLEO provided in Table~\ref{tab:freestream}, $\zeta_{cc}$ is calculated to be 46.4\% for CPM. This indicates that over 50\% of the particles incident to CPM do not reach the cryopanel and are lost due to backflow. Therefore, optimizing the geometrical design to maximize the probability of captured particles impacting the cryopanel while minimizing backflow could enhance $\zeta_{cc}$ and improve the overall intake performance of CRAID.\\ 

\subsubsection{System requirements: $Q_{cool}$ and $P_{cool}$}
\label{subsec:cooler_capacity}

During the condensation sequence, the cryocooler operates to maintain a constant cryopanel temperature. The required cooling capacity can be estimated from the heat transferred to the cryopanel. The heat load on the cryopanels originates from three sources: kinetic energy heat flux, phase change enthalpy, and radiative heat transfer. To determine the cryocooler capacity required for CPM, the kinetic energy heat flux and phase change enthalpy are obtained from DSMC calculations, while the radiative heat transfer is estimated using an analytic relation. 
\begin{figure}[htb!] 
	\centering
	\subfloat[Kinetic energy heat flux $q_k$.]{\includegraphics[width=5.5cm]{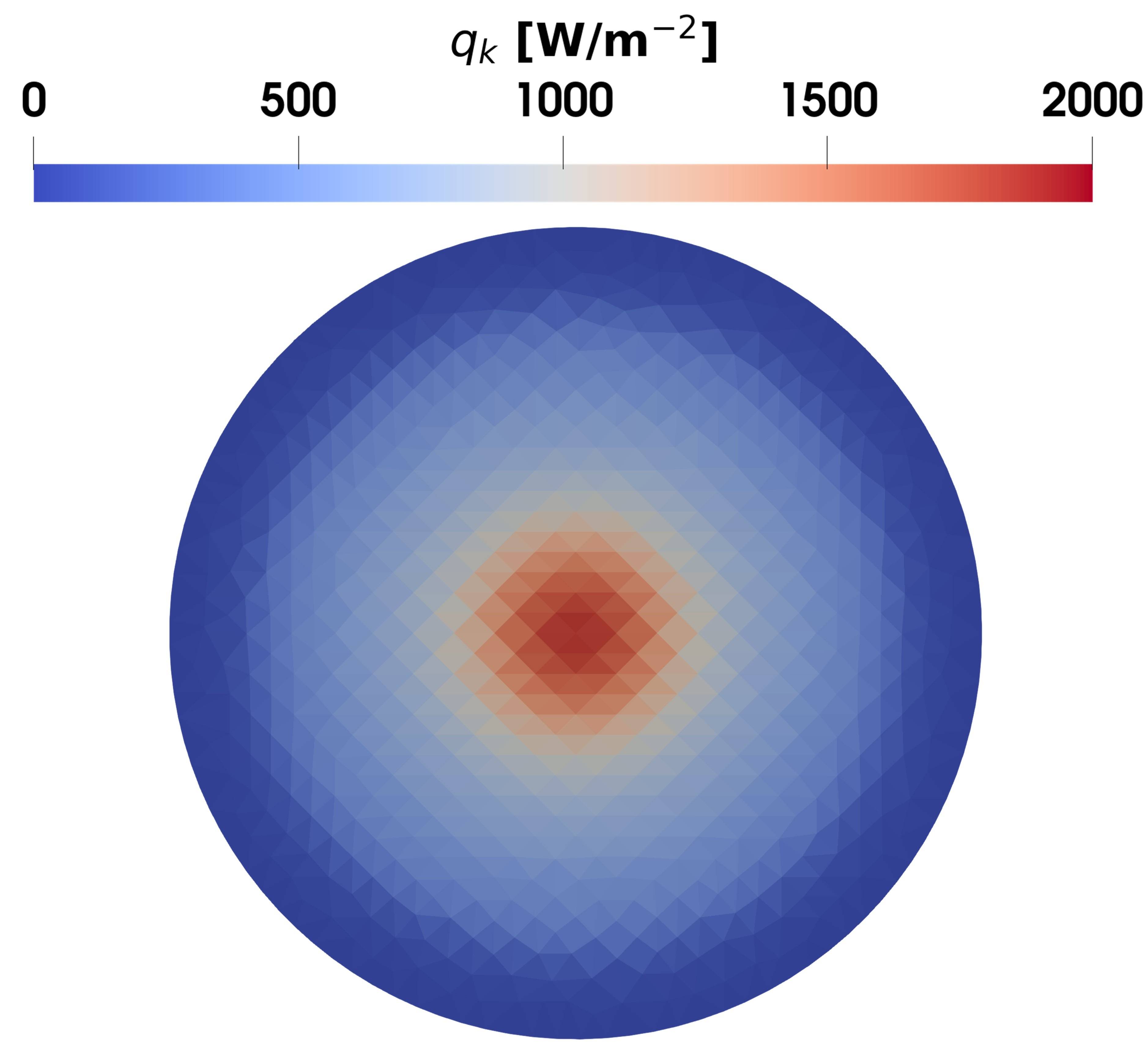}\label{fig:q_kinetic}}
	\hfill
	\subfloat[Phase change enthalpy $q_{ph}$.]{\includegraphics[width=5.5cm]{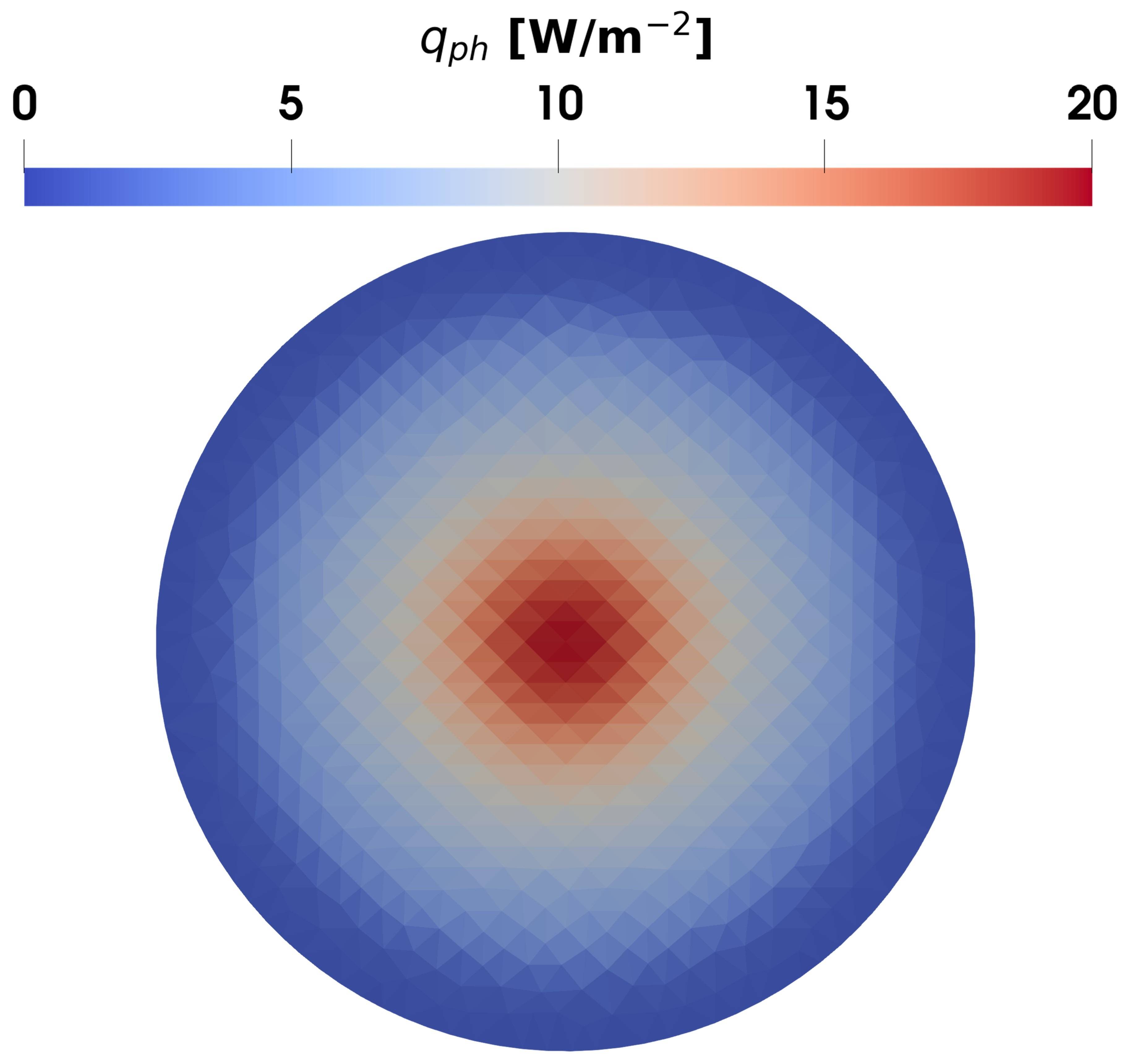}\label{fig:q_phase}}
	\caption{Kinetic energy heat flux and phase change enthalpy on the cryopanel.}
	\label{fig:q_particle}
\end{figure}
First, the kinetic energy heat flux $q_k$ is steadily loaded by incident molecules over time and its local distribution reaches up to 1850 W/m$^{2}$. Fig.~\ref{fig:q_particle}(a) shows the distribution of $q_k$ on the cryopanel. By integrating $q_k$ over $A_{pan}$, the total kinetic energy heat $Q_k$ is calculated to be 13.79 W. Second, the phase change enthalpy $q_{ph}$ is exerted up to 20 W/m$^{2}$ at the center of the cryopanel's frontal surface. The distribution of $q_{ph}$ on the cryopanel is shown in Fig.~\ref{fig:q_particle}(b). The total heat of phase change enthalpy $Q_{ph}$ is calculated to be 0.19 W over $A_{pan}$. Third, the total radiative heat $Q_{rad}$ exerted on the cryopanels can be calculated as~\cite{Nandagopal2022}:
\begin{equation}
    Q_{rad} = \dfrac{\sigma_{SB}(T_{res}^4-T_{pan}^4)}{\dfrac{1-\epsilon_{pan}}{\epsilon_{pan} A_{pan}}+\dfrac{1-\epsilon_{res}}{\epsilon_{res} A_{res}}+\dfrac{1}{A_{res} F_{res\rightarrow pan}}},
    \label{eq:radiative}
\end{equation}
where $\sigma_{SB}$ is the Stefan-Boltzmann constant, $\epsilon$ is the emissivity, $A$ is the surface area, and $F$ is the view factor. The subscripts \textit{res} and \textit{pan} indicate the reservoir and the cryopanel, respectively. Since the cryopanel is enclosed by the reservoir, the reciprocity rule gives $A_{res}F_{res \rightarrow pan} = A_{pan}F_{pan \rightarrow res} = A_{pan}$~\cite{Nandagopal2022}. Considering that both the reservoir and the cryopanel are gray bodies with an emissivity of $\epsilon = 0.03$ for copper~\cite{brewster1992thermal}, $Q_{rad}$ is calculated to be $1.2 \times 10^{-3}$ W. If the oxidized layer on the copper surface is considered, with $\epsilon$ increasing up to 0.9, $Q_{rad}$ remains below 0.03 W~\cite{brewster1992thermal}. In total, approximately 14 W of heat is applied to the cryopanels, corresponding to the $Q_{cool}$ required for CPM. Various active cryocooler mechanisms, including Joule-Thomson expanders, Stirling cycles, and pulse tubes, have been developed and utilized in space missions~\cite{Chen2024}. Among these cryocooling mechanisms, the Reverse Turbo-Brayton (RTB) cycle cryocooler has demonstrated power-efficient cooling performance at 20 K~\cite{Plachta2018, Nugent2022, Nugent2024}. Recent tests of the RTB cryocooler have demonstrated a cooling capacity of 20 W at 20 K with an input power of 1.7 kW, yielding a specific power of 85 W/W~\cite{Nugent2024}. Therefore, an RTB cryocooler with a power consumption $P_{cool}$ of 1.2 kW would be required to provide a $Q_{cool}$ of 14 W during the condensation sequence.\\


\subsection{Regeneration sequence}

\subsubsection{Gas transport characteristics}
After the condensation sequence simulation is completed, the recorded condensed particle density distribution on the cryopanel is transferred to a separate simulation for the regeneration sequence. The cryopanels' temperature is raised to 54.5 K. The simulation domain is modified to include the double propellant injection tubes, while excluding the passive intake device. The domain size is reduced to align with the reservoir dimensions as shown in Fig.~\ref{fig:domain_regen}. The domain is discretized into cubic calculation grid cells using a 5-level octree AMR technique. The time step size is set to $10^{-6}$ seconds to resolve the mean collision time of the regenerated gas, and the macroscopic properties of the flow field are tallied every 100,000 steps.\\

\begin{figure}[htb!] 
	\centering
	\includegraphics[width=12cm]{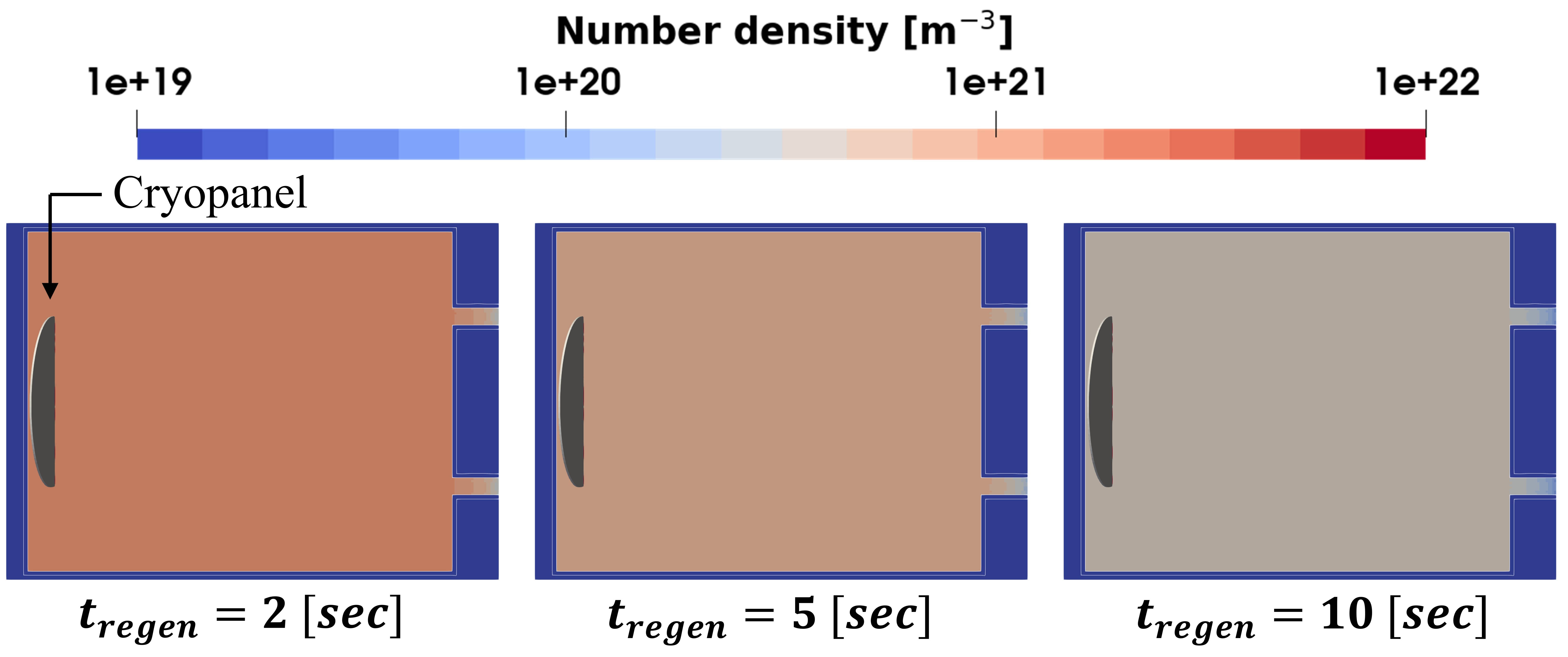}
	\caption{Number density contours during the regeneration sequence.}
	\label{fig:regen_contour}
\end{figure}

Gas transport during the regeneration sequence is investigated using the condensed particle density recorded at $t_{cond}=10$ seconds, as obtained in Sec.~\ref{subsec:cond1}. DSMC calculations are conducted for the same duration of $t_{regen}=10$ seconds, and the number density contours of regenerated gas molecules over time are presented in Fig.~\ref{fig:regen_contour}.
The number density of the regenerated gas particles gradually decreases over time as they flow through the propellant injection tubes. The transport properties of the regenerated gas are characterized by the instantaneous capture efficiency $\eta_c^I$ and compression ratio $CR^I$~\cite{Moon2024ast}:
\begin{equation}
	\eta_c^I(t)=\frac{\dot{N}_T(t)}{\dot{N}_\infty}=\frac{\bar{n}_{T}(t)\bar{v}_{T}(t)A_{T}}{n_\infty v_\infty A_{in}},
	\label{eq:eta_inst}
\end{equation}
\begin{equation}
	CR^I(t) = \frac{\bar{n}_{T}(t)}{n_\infty},
	\label{eq:CR_inst}
\end{equation}
which are analogous to the conventional $\eta_c$ and $CR$, but account for the particle flow rate $\dot{N}_T(t)$ and number density $n_T(t)$ through the propellant injection tube at time $t$, relative to the freestream conditions. Here, $\bar{n}_T$ and $\bar{v}_T$ represent the mean number density and velocity at the entrance area $A_T$ of the propellant injection tube, respectively.
\begin{figure}[htb!] 
	\centering
	\subfloat[$\eta_c^I$.]{\includegraphics[width=8.5cm]{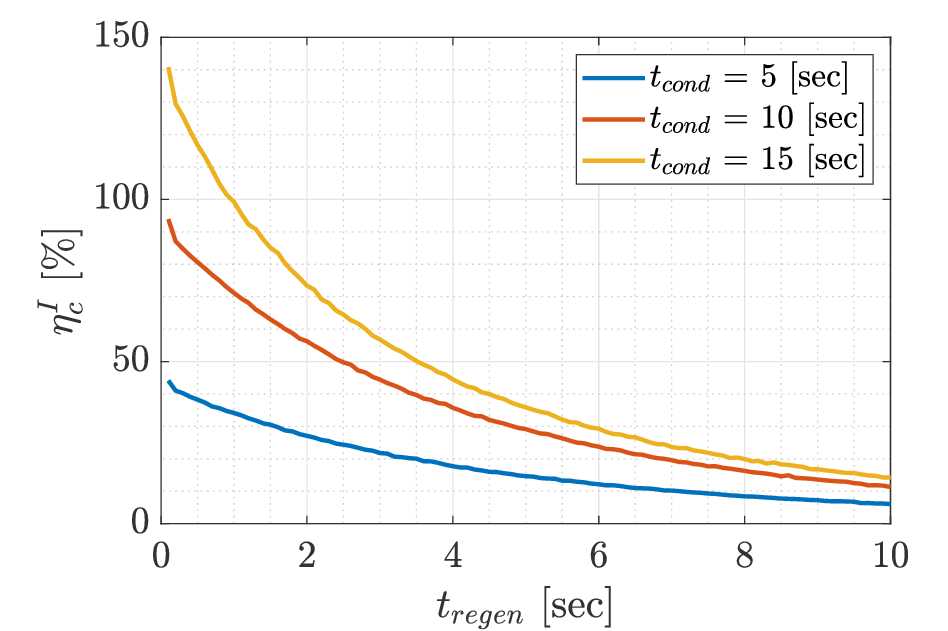}\label{fig:eta_I}}
	\vfill
	\subfloat[$CR^I$.]{\includegraphics[width=8.5cm]{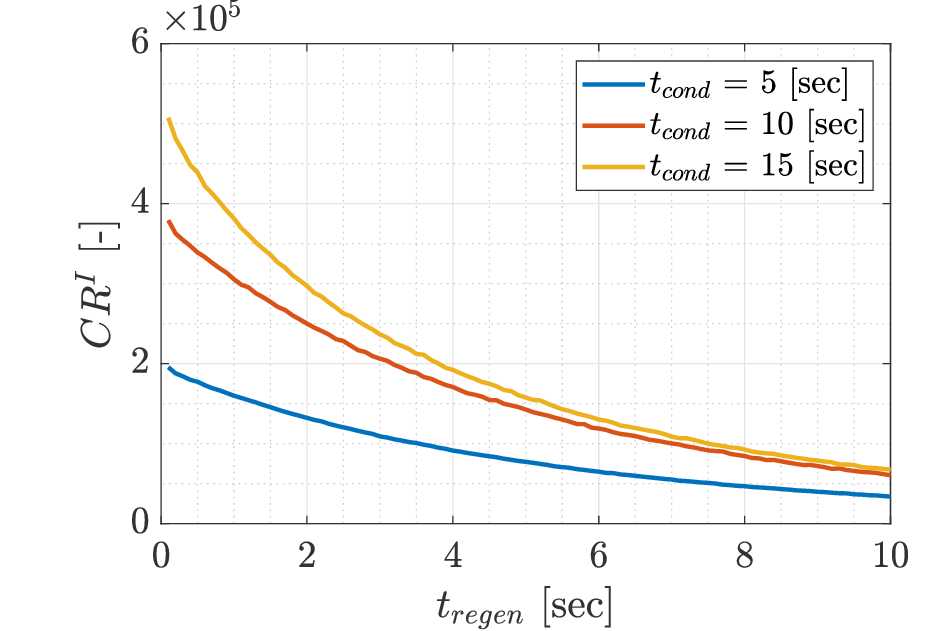}\label{fig:CR_I}}
	\caption{Instantaneous intake performance for various $t_{cond}$.}
	\label{fig:perf_I}
\end{figure}
The DSMC calculation for the regeneration sequence is additionally performed using condensed particles recorded at $t_{cond}=5$ and 15 seconds to characterize the variation in instantaneous performance when $t_{cond}$ is reduced or extended. Fig.~\ref{fig:perf_I} illustrates the instantaneous intake performance with respect to $t_{regen}$. In all cases, the regenerated gas rapidly fills the reservoir at the beginning of the regeneration sequence. Both $\eta_c^I$ and $CR^I$ exhibit exponential decay over time because the flow rate escaping the reservoir, $-\dot{N_T}$, is proportional to the density of the transferring regenerated gas, $\bar{n}_T\approx N_T/V_{res}$. For the simulation case with $t_{cond} = 10$ seconds, $\eta_c^I$ decays from 94\% to 11\% and $CR^I$ drops from $3.8 \times 10^5$ to $6.1 \times 10^4$ over a $t_{regen}$ of 10 seconds. When $t_{cond}$ is extended to 15 seconds, the maximum $\eta_c^I$ increases to 141\%, whereas reducing $t_{cond}$ to 5 seconds decreases the maximum $\eta_c^I$ to 44\%. The maximum $CR^I$ increases from $2.0 \times 10^5$ to $5.1 \times 10^5$ as $t_{cond}$ is extended from 5 to 15 seconds. This suggests that extending $t_{cond}$ can improve instantaneous intake performance during the regeneration sequence.\\

\subsubsection{System requirements: $t_{cond}^*$ and $t_{regen}^*$}

\begin{figure}[htb!] 
	\centering
	\subfloat[Partial pressure.]{\includegraphics[width=8.5cm]{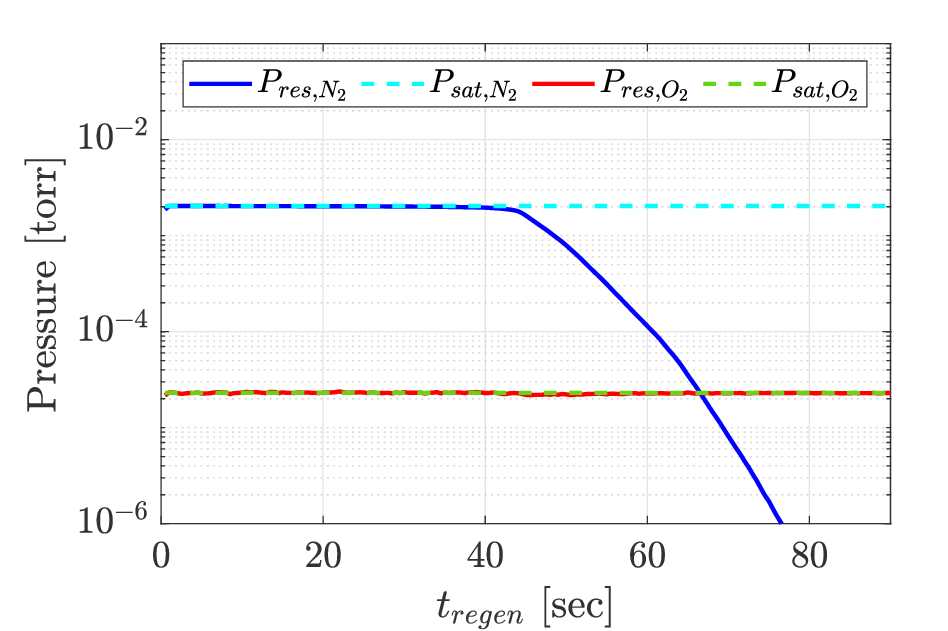}\label{fig:P_res}}
	\vfill
	\subfloat[Chemical composition.]{\includegraphics[width=8.5cm]{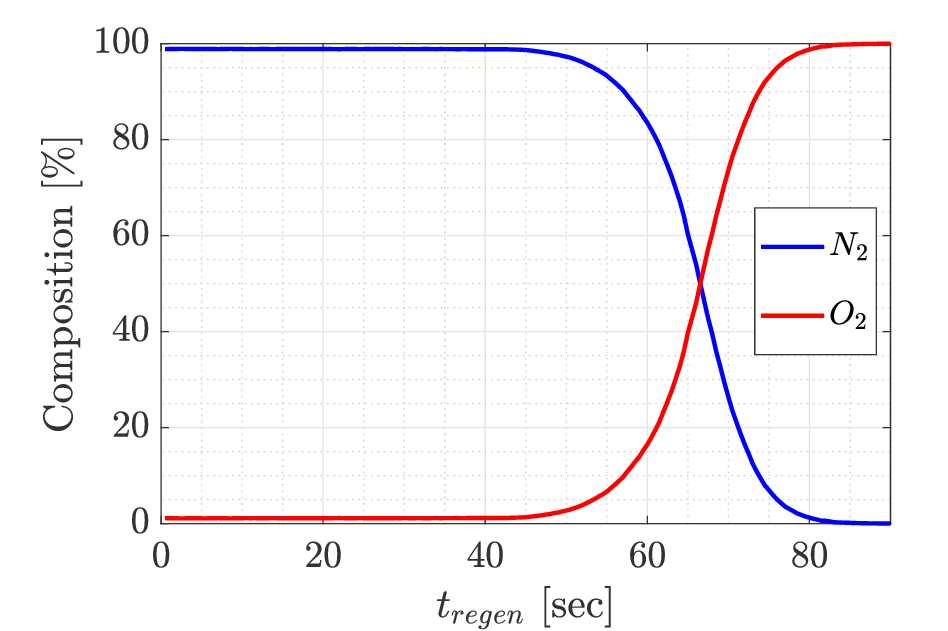}\label{fig:com_res}}
	\caption{Partial pressure and chemical composition of regenerated gas in the reservoir during the saturated operation at 35 K.}
	\label{fig:P_res_sat}
\end{figure}

Extending $t_{cond}$ can be beneficial for instantaneous intake performance; however, its upper limit is constrained by the saturation of regenerated gas. As $t_{cond}$ increases, the initial pressure of the regenerated gas increases and may reach the saturated vapor pressure. The CRAID operation with saturated gas regeneration is termed a saturated operation, whereas the case without saturation is referred to as an under-saturated operation. If CRAID reaches saturated operation, species-selective propellant transfer may occur due to differences in the saturated vapor pressures of each species. Given that $N_2$ exhibits at least 10 times higher saturated vapor pressure than $O_2$ over 20--54.5 K, the species-selective effect can be observed regardless of the cryopanel temperature during the regeneration sequence~\cite{brown1980vapor}. An additional simulation is conducted to investigate the species-selective behavior of propellant transport during saturated operation, applying condensed particle density from an extended $t_{cond}$ of 30 seconds, with the cryopanel temperature lowered to 35 K to ensure saturation during gas regeneration. The simulation for saturated operation is performed for $t_{regen}=90$ seconds, which provides sufficient duration to observe the species-selective effect. Fig.~\ref{fig:P_res_sat} shows the calculated partial pressure and the chemical composition of the regenerated gas inside the reservoir.
Before $t_{regen} = 45$ seconds in Fig.~\ref{fig:P_res}, the regenerated $N_2$ and $O_2$ molecules maintain constant partial pressures, which correspond to their respective saturated vapor pressures at 35 K. Since the regenerated $N_2$ gas exhibits approximately 87 times higher pressure than $O_2$, $N_2$ gas escapes the reservoir much faster. This results in earlier depletion of $N_2$ gas, with a rapid decrease of its partial pressure by $t_{regen}=45$ seconds. The gas transferred through the propellant injection tube initially contains 99\% $N_2$, as shown in Fig.~\ref{fig:com_res}. However, after a transition period between $t_{cond}$ = 45 and 80 seconds, the composition shifts to 100\% $O_2$. A similar species-selective effect can be observed during saturated operation at various cryopanel temperatures, including the 54.5 K used in CPM. Due to the species-selective effect in saturated operation, limiting $t_{cond}$ to maintain under-saturated operation may be necessary for the stable operation of ABEP thrusters. The maximum $t_{cond}$ for under-saturated operation should be determined by the regenerated $O_2$ gas pressure, as the saturated vapor pressure of $O_2$ is approximately ten times lower than that of $N_2$ at 54.5 K. Assuming a 50\% under-saturated operation, $t_{cond,50\%}$ is given by:
\begin{equation}
    t_{cond,50\%} = 0.5\times \frac{n_{sat,O_2} V_{res}}{\zeta_{cc} n_{\infty,O_2} u_\infty A_{in}}.
    \label{eq:t_cond_50}
\end{equation}
Here, $n_{sat,O_2}$ represents the number density of $O_2$ at saturation, given as $2.0 \times 10^{23}$ m$^{-3}$ at 54.5 K. Given that CPM has a $V_{res}$ of 0.12 m$^{3}$ and the freestream conditions listed in Table~\ref{tab:freestream}, $t_{cond,50\%}$ is calculated to be 42.4 minutes. Therefore, a practical condensation sequence duration $t_{cond}^*$ is determined to be 42.4 minutes at 200 km altitude, and it varies with altitude as the freestream particle flux changes.\\ 
%

\begin{figure}[htb!]
	\centering
	\includegraphics[width=12cm]{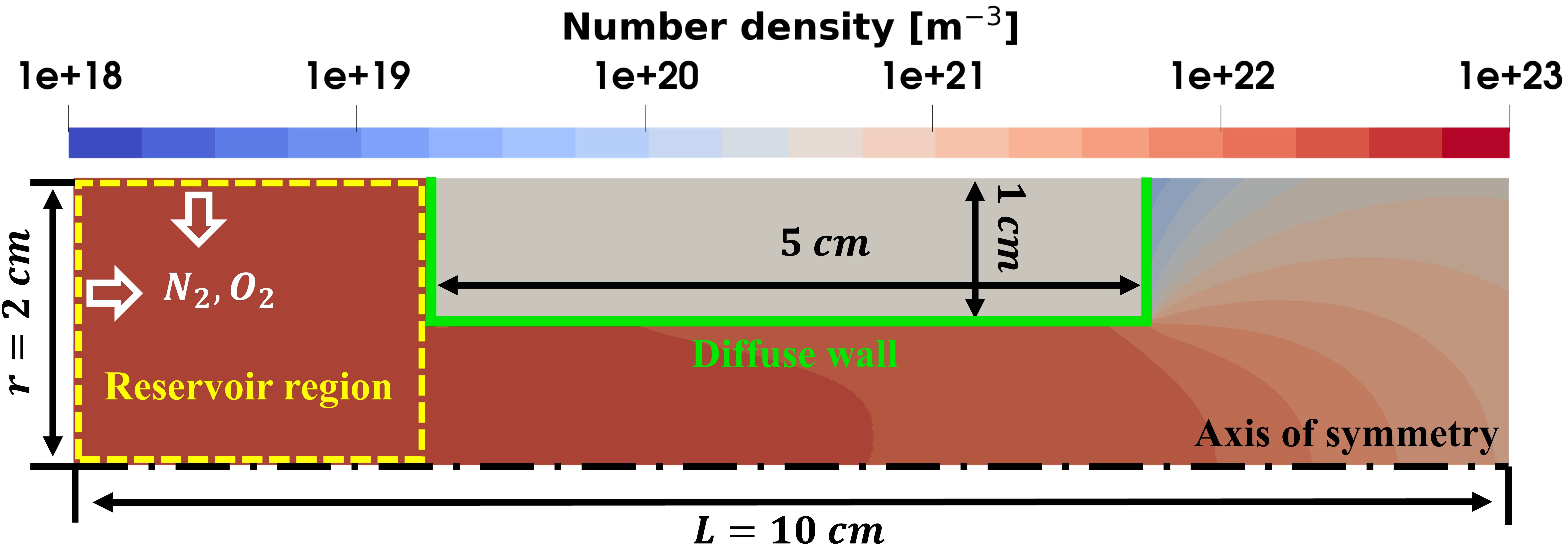}
	\caption{Simulation domain and example number density contours from axisymmetric circular channel calculation.}
	\label{fig:tube_domain}
\end{figure}

\begin{figure}[htb!]
	\centering
	\subfloat[Linear scale.]{\includegraphics[width=9.5cm]{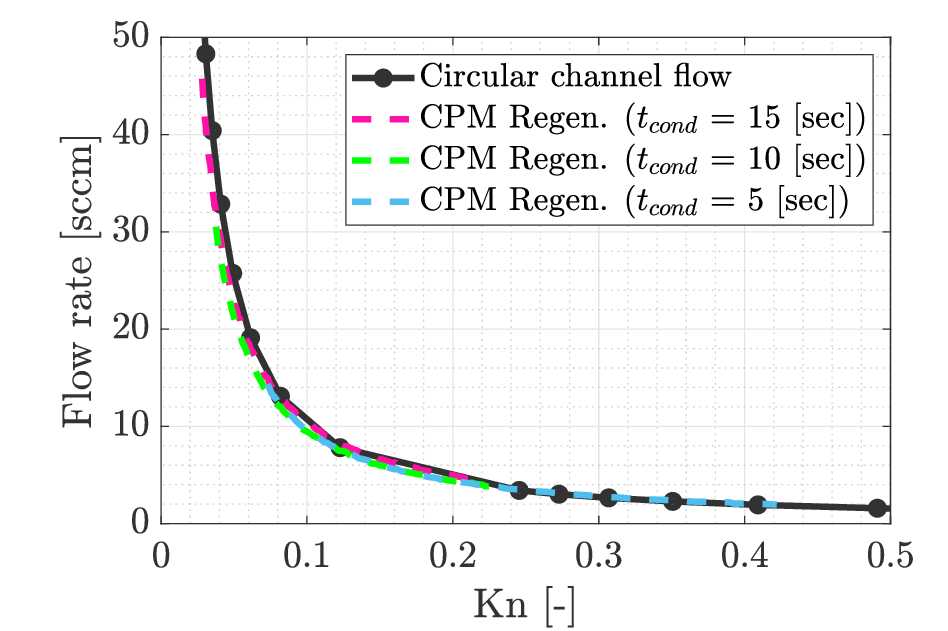}\label{fig:kn_lin}}
	\vfill
	\subfloat[Log scale.]{\includegraphics[width=9.5cm]{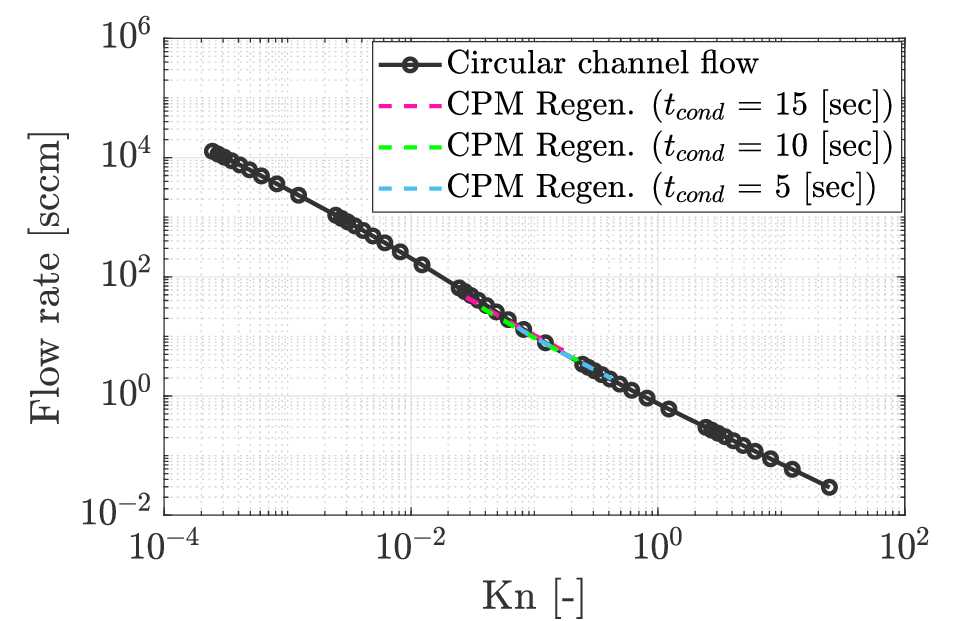}\label{fig:kn_log}}
	\caption{$Kn$ -- flow rate characterization.}
	\label{fig:tube_plot}
\end{figure}

\begin{figure}[htb!] 
	\centering
	\subfloat[Linear scale.]{\includegraphics[width=8.5cm]{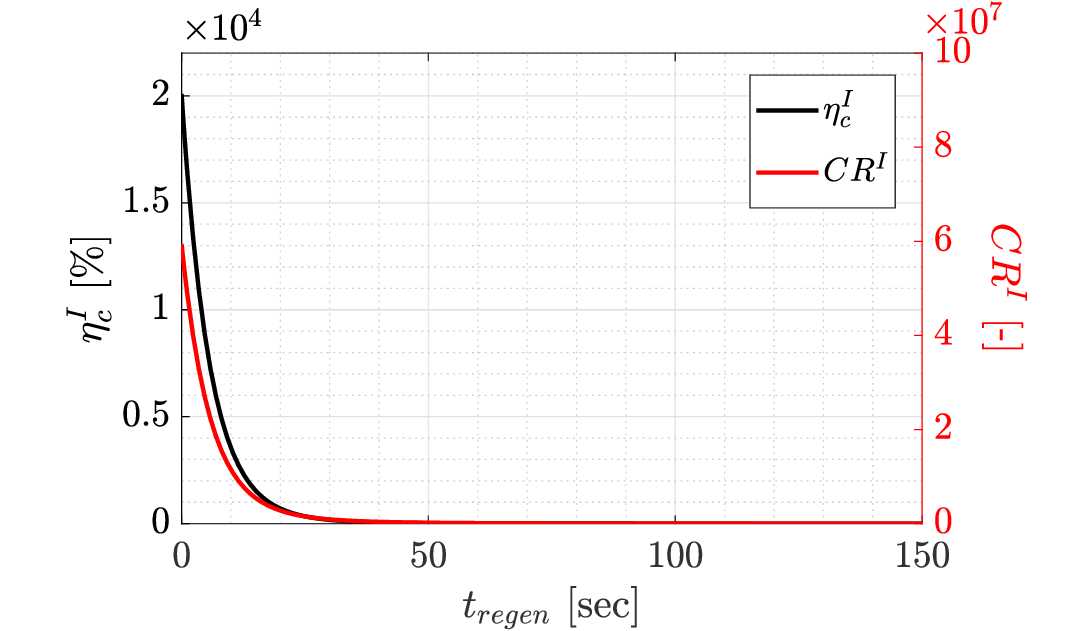}\label{fig:noMFC_lin}}
	\vfill
	\subfloat[Log scale.]{\includegraphics[width=8.5cm]{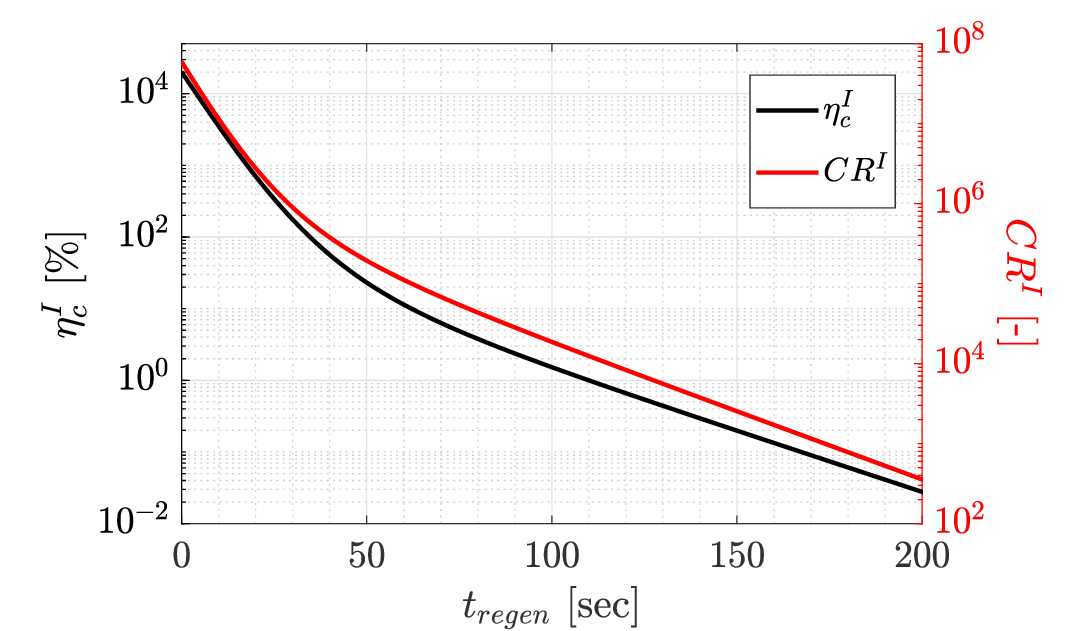}\label{fig:noMFC_log}}
	\caption{Instantaneous intake performance for 50\% under-saturated operation.}
	\label{fig:noMFC}
\end{figure}

A practical $t_{regen}^*$ for utilizing the gas condensed during $t_{cond}^*$ is determined by analyzing the transport characteristics of the regenerated gas. As shown in Fig.~$\ref{fig:perf_I}$, the condensed particles rapidly regenerate and fill the reservoir at the beginning of the regeneration sequence during under-saturated operation. The transport of the regenerated gas is driven by the pressure difference between the two ends of the propellant injection tube. Therefore, the gas transport during under-saturated operation can be approximated as circular channel flow. The transport of the regenerated gas inside CPM is analyzed across a wide range of pressure difference using DSMC calculations for circular channel flow. The circular channel has a radius $r_{T}$ of 1 cm and a length of 5 cm, corresponding to the dimensions of the propellant injection tubes in CPM. DSMC calculations are performed in a 2D axisymmetric domain, as illustrated in Fig.~\ref{fig:tube_domain}. In the simulation domain, the left side of the channel represents the region inside the reservoir, where $N_2$ and $O_2$ molecules are injected from the domain boundaries, except along the axis of symmetry. The flow rate through the circular channel is characterized as a function of the reservoir's degree of rarefaction, by varying the injected $O_2$ number density across 46 cases, ranging from $10^{18}$ to $10^{23}$ m$^{-3}$ at a constant temperature of 54.5 K. This range encompasses the $O_2$ number densities achievable in the reservoir during 50\% under-saturated operation. In proportion to the $O_2$ number density, $N_2$ molecules are also injected, to maintain a chemical composition consistent with the freestream conditions. The total number density of the generated gas ranges from $2.8\times 10^{18}$ to $2.8\times 10^{23}$. The circular channel walls are assigned a fully diffuse GSI model with a surface temperature of 54.5 K. A rectangular structured grid is employed with a 5-level quad-tree AMR technique to adjust cell sizes so that the molecular mean free path is resolved in each case. The time step is set to be shorter than the mean collision time. The particle flow rate through the circular channel is obtained once the simulations reach a steady state.
The calculated flow rate through the circular channel is correlated with the $Kn$ of the transferred gas, as shown in Fig.~\ref{fig:tube_plot}. The $Kn$ is evaluated at the channel entrance based on the gas mean free path, with the channel diameter as the characteristic length. The black symbolic line represents the profile obtained from the circular channel flow calculation. The dashed lines represent the profiles obtained from the regeneration sequence simulations of CPM with $t_{cond}=$5, 10, and 15 seconds. The $Kn$--flow rate profile from the circular channel flow calculation agrees with CPM regeneration sequence simulations within 5\%, indicating that the circular channel flow can accurately characterize the flow through the propellant injection tube of CPM. With the extensively characterized $Kn$--flow rate relation, the flow rate $\dot{N}_T(t)$ and number density $\bar{n}_T(t)$ during the regeneration sequence can be expressed as:
\begin{equation}
	\dot{N}_T(t) = -f(Kn(t)) = -g(\bar{n}_T(t)),
	\label{eq:dNdt}
\end{equation}
\begin{equation}
	\bar{n}_T(t) = -\int_{0}^{t_{regen}}\frac{1}{V_{res}}\dot{N}_T(t)dt.
	\label{eq:nTdt}
\end{equation}
Here, $f(Kn)$ is the interpolated function from Fig.~\ref{fig:tube_plot}. By integrating Eqs. (\ref{eq:dNdt}) and (\ref{eq:nTdt}), $\dot{N}_T(t)$ and $\bar{n}_T(t)$ are calculated as a function of elapsed time during the regeneration sequence. An initial condition of $\bar{n}_T(0) = 0.5 \times (n_{sat,O_2} + n_{sat,N_2}) = 2.8 \times 10^{23}$ is applied to the integration, representing the initial number density for 50\% under-saturated operation. Normalizing $\dot{N}_T(t)$ and $\bar{n}_T(t)$ with the freestream conditions defined in Eqs. (\ref{eq:eta_inst}) and (\ref{eq:CR_inst}), the instantaneous intake performances for 50\% under-saturated operation are estimated over a $t_{regen}$ of 200 seconds and shown in Fig.~\ref{fig:noMFC}. At $t_{regen}=$200 seconds, the gas transport becomes negligible with $\eta_c^I$ of less than 0.3\%. At the beginning of the regeneration sequence, $\eta_c^I$ and $CR^I$ reach $2 \times 10^4$\% and $6 \times 10^7$, respectively, which are more than 100 times higher than the instantaneous intake performance observed in Fig.~\ref{fig:perf_I} for $t_{cond}=15$ seconds. However, the performance curves rapidly decay, falling below 1/1000 of its initial values within 50 seconds. This drastic change in propellant injection properties disrupts the discharge in EP thrusters. Consequently, a practical $t_{regen}^*$ cannot be determined unless the propellant flow rate is regulated to maintain steady thruster operation.\\ 

\begin{figure}[htb!] 
	\centering
	\subfloat[$\eta_c^I$.]{\includegraphics[width=8.5cm]{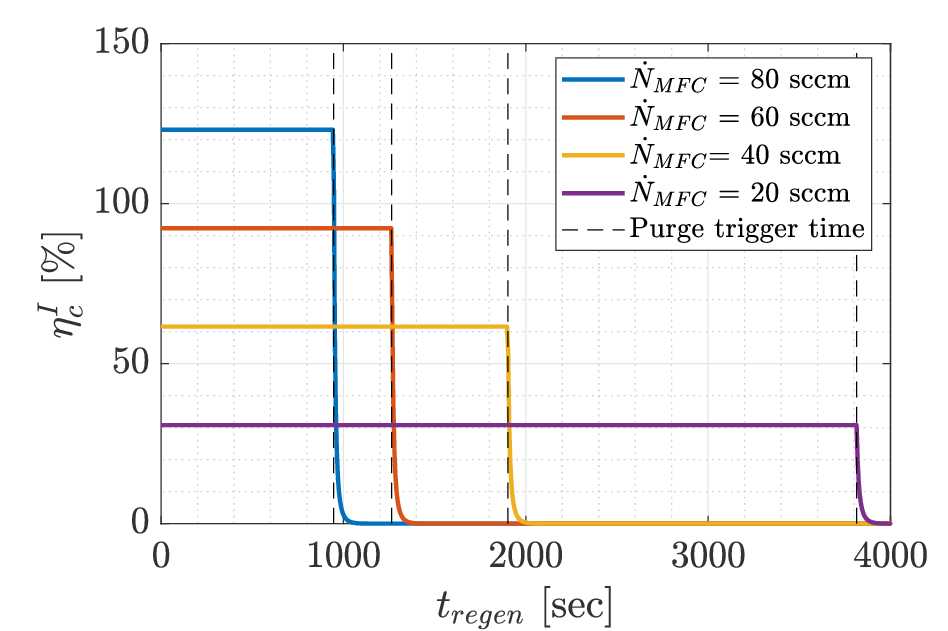}\label{fig:MFC_eta}}
	\vfill
	\subfloat[$CR^I$.]{\includegraphics[width=8.5cm]{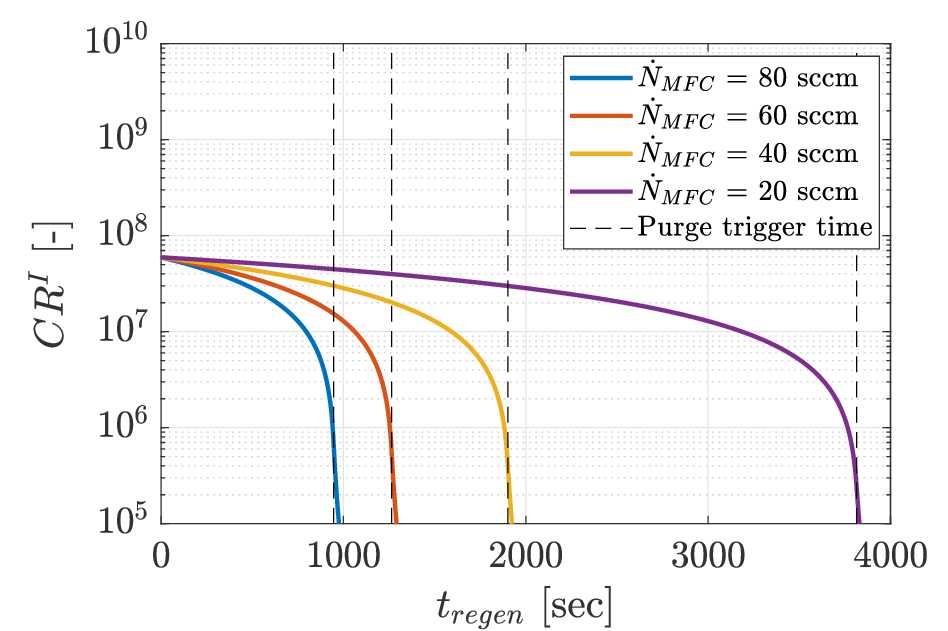}\label{fig:MFC_CR}}
	\caption{Instantaneous intake performance with MFC applied for 50\% under-saturated operation.}
	\label{fig:MFC_perf}
\end{figure}

The MFC is integrated into CRAID system to regulate the propellant flow rate. As a target flow rate $\dot{N}_{MFC}$ is specified, the MFC actuates a proportional valve to mechanically adjust the cross-sectional area of the flow path to achieve the desired flow rate. Considering various $\dot{N}_{MFC}$ for the MFC, the instantaneous intake performance of CPM during 50\% under-saturated operation is recalculated by modifying Eq.~(\ref{eq:dNdt}) as:
\begin{equation}
	\dot{N}_T(t)=\begin{cases}
    -\dot{N}_{MFC}, & \text{if } \dot{N}_{MFC} \le g(\bar{n}_T(t)), \\
    -g(\bar{n}_T(t)), & \text{if } \dot{N}_{MFC}> g(\bar{n}_T(t)).
    \end{cases}
	\label{eq:dNdt_mod}
\end{equation}
This equation indicates that when the MFC is active, narrowing the proportional valve restricts the flow rate to $\dot{N}_{MFC}$. However, if $\dot{N}_{MFC}$ cannot be achieved even when the valves are fully open, allowing flow through the entire propellant injection tube, the flow rate is determined without MFC regulation. By integrating Eqs.~(\ref{eq:dNdt_mod}) and (\ref{eq:nTdt}), and normalizing the results following Eqs.~(\ref{eq:eta_inst}) and (\ref{eq:CR_inst}), the instantaneous intake performances with MFC are obtained as shown in Fig.~\ref{fig:MFC_perf}. Since $\eta_c^I$ represents the propellant flow rate relative to the freestream, it remains constant at the targeted $\dot{N}_{MFC}$ as shown in Fig~\ref{fig:MFC_eta}. When most of the regenerated gas in the reservoir is depleted, the gas flow through the entire cross-sectional area of the propellant injection tube can no longer sustain $\dot{N}_{MFC}$. This point is referred to as the purge trigger time. As $\dot{N}_{MFC}$ decreases from 80 sccm to 20 sccm, the purge trigger time is delayed from 940 seconds to 3810 seconds because of the slower depletion of regenerated gas. After the purge trigger time, the thrusters are assumed to cease operation, and the remaining gas in the reservoir is rapidly exhausted within tens of seconds. The purge trigger time is considered the effective duration of the thruster operation, corresponding to $t_{regen}^*$. Assuming a propellant flow rate of 20 sccm for two RITs, the practical $t_{regen}^*$ for CPM is estimated to be 31.7 minutes. In Fig.~\ref{fig:MFC_CR}, $CR^I$ gradually decreases from $5.9\times 10^7$ but remains above $10^5$ until the purge trigger time. This value surpasses the minimum $CR$ threshold required for sustaining discharge in conventional EP thrusters. Therefore, integrating the MFC with CRAID ensures a steady propellant supply to the thruster, allowing for stable thrust generation over an extended duration.\\ 

\subsection{Cool-down sequence}

The practical $t_{cool}^*$ for the cool-down sequence can be estimated based on the heat capacity of the cryopanel and the cooling capacity of the cryocooler. The cryopanel of CPM is assumed to be made of oxygen-free high conductivity copper (OFHC) which exhibits a low heat capacity at cryogenic temperatures. Using the OFHC density of 0.0090 g/mm$^{3}$, the cryopanel mass $m_{pan}$ in CPM is calculated as 0.56 kg. The specific heat $h_{sp}$ of OFHC at temperature $T$ is given by the empirical equation~\cite{marquardt2002cryogenic}:
\begin{multline}
    \log (h_{sp}) = a_0 + a_1\log T + a_2(\log T)^2 + a_3(\log T)^3 \\
    + a_4(\log T)^4 + a_5(\log T)^5 + a_6(\log T)^6 + a_7(\log T)^7,
    \label{eq:h_sp}
\end{multline}
where $a_0$--$a_7$ are the empirical coefficients listed in Table~\ref{tab:heat_coeff}. The total heat $Q_{cool}$ required for the cool-down sequence to reduce $T_{pan}$ from 54.5 K to 20 K is obtained by integrating Eq.~(\ref{eq:h_sp}):
\begin{equation}
    C_{cool} = -\int_{54.5}^{20} m_{pan} h_{sp} dT = 1.01\times 10^3\quad \textrm{[J]}.
    \label{eq:Q_cool}
\end{equation}
Given that the cooling capacity of CPM is determined as $Q_{cool}=14$ W, the practical cool-down sequence duration $t_{cool}^*$ is estimated to be 1.2 minutes. 

\begin{table}
    \centering
    \begin{center}
        \caption{Coefficients for the empirical specific heat of OFHC~\cite{marquardt2002cryogenic}}

        \begin{tabular}{cccccccc}
        \toprule
        $a_0$ & $a_1$ & $a_2$ & $a_3$ & $a_4$ & $a_5$ & $a_6$ & $a_7$ \\ \midrule
        -1.918 & -0.160 & 8.610 & -18.996 & 21.966 & -12.733 & 3.543 & -0.380 \\
        \bottomrule 
        \end{tabular}
        \label{tab:heat_coeff}
    \end{center}
\end{table}

\subsection{Comprehensive system requirement}

\subsubsection{Overall intake performance and system requirements}

The system requirements of CPM, including the practical duration of each sequence and cryocooler specifications, are summarized in Table~\ref{tab:time_overall}, assuming 50\% under-saturated operation of CPM at an altitude of 200 km. The overall intake performance of CRAID is evaluated using the effective compression ratio $CR^{eff}$ and the effective capture efficiency $\eta_c^{eff}$~\cite{Moon2024ast}: 
\begin{equation}
	\eta_c^{eff} = \frac{1}{t_{total}} \int^{t_{regen}}_{0} \eta_c^I (t)dt,
	\label{eq:eta_eff}
\end{equation}
\begin{equation}
	CR^{eff} = \frac{1}{t_{regen}} \int^{t_{regen}}_{0} CR^I (t)dt.
	\label{eq:CR_eff}
\end{equation}
Here, $\eta_c^{eff}$ represents the fraction of the freestream captured and utilized by CRAID over the entire ABEP operational cycle. It provides a comparable metric to the conventional $\eta_c$ used for passive intake devices. On the other hand, $CR^{eff}$ denotes the time-averaged compression ratio during the regeneration sequence, which corresponds to the net duration of the thruster operation. Based on the instantaneous intake performance results shown in Fig.\ref{fig:MFC_perf}, $\eta_c^{eff}$ is determined to be 26.1\% and $CR^{eff}$ is $3.0 \times 10^7$ for CPM, as listed in Table~\ref{tab:time_overall}. A previous study demonstrated that an RIT-based ABEP spacecraft can achieve complete drag compensation near 200 km altitude when $\eta_c$ exceeds 20\%~\cite{Ko2023}. The overall intake performance of CPM suggests that CRAID might achieve complete drag compensation. Furthermore, the significant increase in $CR^{eff}$ compared to conventional passive intake devices offers advantages in maintaining sustained discharge in EP thrusters.

\begin{table}
    \centering
    \begin{center}
        \caption{System requirements and overall intake performance for 50\% under-saturated operation at 200 km altitude.}

        \begin{tabular}{lll}
        \toprule
        \multicolumn{2}{c}{Properties} & Values \\
        \midrule
        \multirow{6}{*}{System requirements}& $t_{total}^*$ & 75.3 [min] \\
        & $t_{cond}^*$ & 42.4 [min] \\
        & $t_{regen}^*$ & 31.7 [min] \\ 
        & $t_{cool}^*$ & 1.2 [min] \\
        & $Q_{cool}$ & 14 [W] \\
        & $P_{cool}$ & 1200 [W] \\
        \midrule
        \multirow{2}{*}{Overall intake performance} & $\eta_c^{eff}$ & 26.1\% \\
        & $CR^{eff}$ &  $3.0\times 10^7$ \\ 
        \bottomrule
        \end{tabular}
        \label{tab:time_overall}
    \end{center}
\end{table}


\subsubsection{Variation of system requirements at different altitudes}
\begin{figure}[htb!]
	\centering
	\includegraphics[width=10cm]{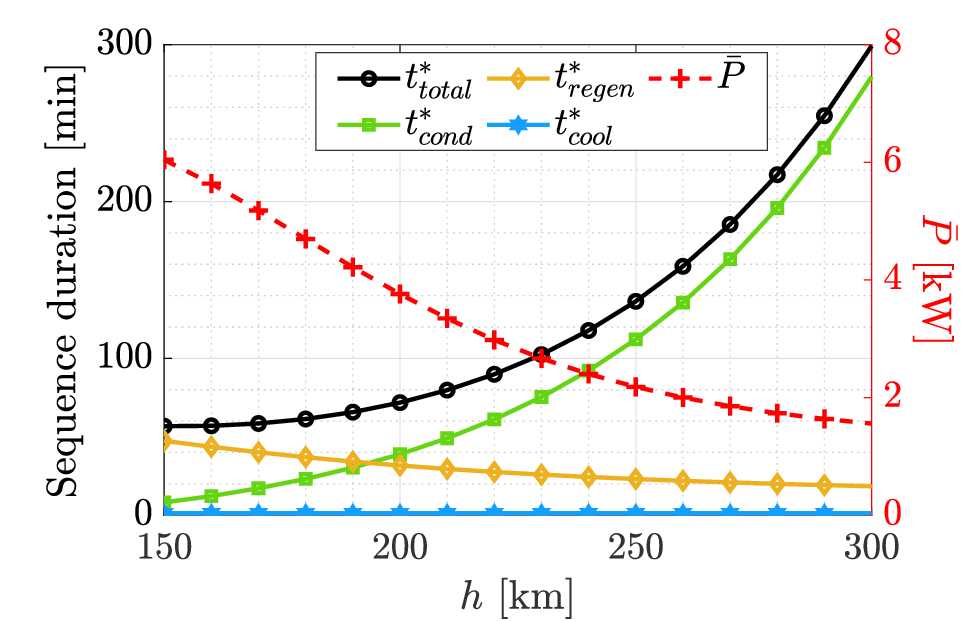}
	\caption{Estimated sequence duration and average power consumption $\bar{P}$ during an operational cycle of CRAID.}
	\label{fig:seq_time_power}
\end{figure}

The practical sequence durations for CRAID operation vary with altitude because the incident freestream flux depends on atmospheric density. The relative allocation of each operational sequence within $t_{total}^*$ is crucial for assessing CRAID's overall intake performance and drag compensation capability. Variations in $t_{cond}^*$, $t_{regen}^*$, and $t_{cool}^*$ are investigated at altitudes ranging from 150 km to 300 km in 10 km intervals. At each altitude, $t_{cond}^*$ is determined using Eq.~(\ref{eq:t_cond_50}) to satisfy 50\% under-saturated operation, and $t_{regen}^*$ is assigned to maintain a propellant flow rate of 20 sccm for each of the two RITs. Since the cool-down sequence is independent of the environment, $t_{cool}^*$ remains constant at 1.2 minutes regardless of altitude. The freestream conditions are derived from the NRLMSIS 2.0 model and are global-averaged under normal solar activity~\cite{Emmert2020}. The estimated sequence durations at different altitudes are depicted in Fig.~\ref{fig:seq_time_power} as symbolic solid lines. The estimated $t_{cond}$ exhibits a monotonic increase from 8.2 to 279.5 minutes as altitude increases due to the decrease in atmospheric density. As altitude increases, the proportion of oxygen in the atmosphere increases, while that of nitrogen decreases. This reduces the total amount of gas condensed at a given level of oxygen saturation. As a result, the $t_{regen}$ for 50\% under-saturated operation with a $\dot{N}_{MFC}=20$ sccm decreases from 47.4 to 18.5 minutes as altitude increases. Since $t_{cond}^*$ accounts for more than 50\% of the total operational cycle, $t_{total}^*$ increases from 56.7 to 299.1 minutes as altitude increases. Variations in the time allocation for each operational sequence with altitude indicate that the duration available for thrust production varies. To ensure the stable operation of satellites equipped with CRAID, it is essential to analyze orbital correction maneuver scenarios considering these variations.\\


Proper power budgeting is also crucial for the design of a CRAID-integrated satellite. The average power consumption $\bar{P}$ over the entire operational cycle is calculated to evaluate the power budget required for CPM. During the condensation and cool-down sequences, the thrusters remain inactive, while $P_{cool}$ of 1.2 kW is required to operate the cryocooler. Conversely, during the regeneration sequence, the cryocooler is inactive, while the thrusters operate. Assuming CRAID is equipped with 3.5 kW-class twin RITs, $\bar{P}$ is calculated as:
\begin{equation}
    \bar{P} = \frac{1.2 (t_{cond}^*+t_{cool}^*) + 7 t_{regen}^*}{t_{total}^*} \quad [\textrm{kW}].
\end{equation}
The red dashed symbolic line in Fig.~\ref{fig:seq_time_power} represents the calculated $\bar{P}$ at various altitudes. At higher altitudes, the $t_{regen}^*$ for thrust generation constitutes a smaller fraction of the total operational cycle. Since the thrusters require high power, the power budget for CPM operation decreases from 6.0 kW to 1.6 kW as altitude increases.  

\subsection{Flight envelope of CRAID-integrated ABEP}

\subsubsection{Satellite modeling}
\begin{figure}[htb!] 
	\centering
	\includegraphics[width=10cm]{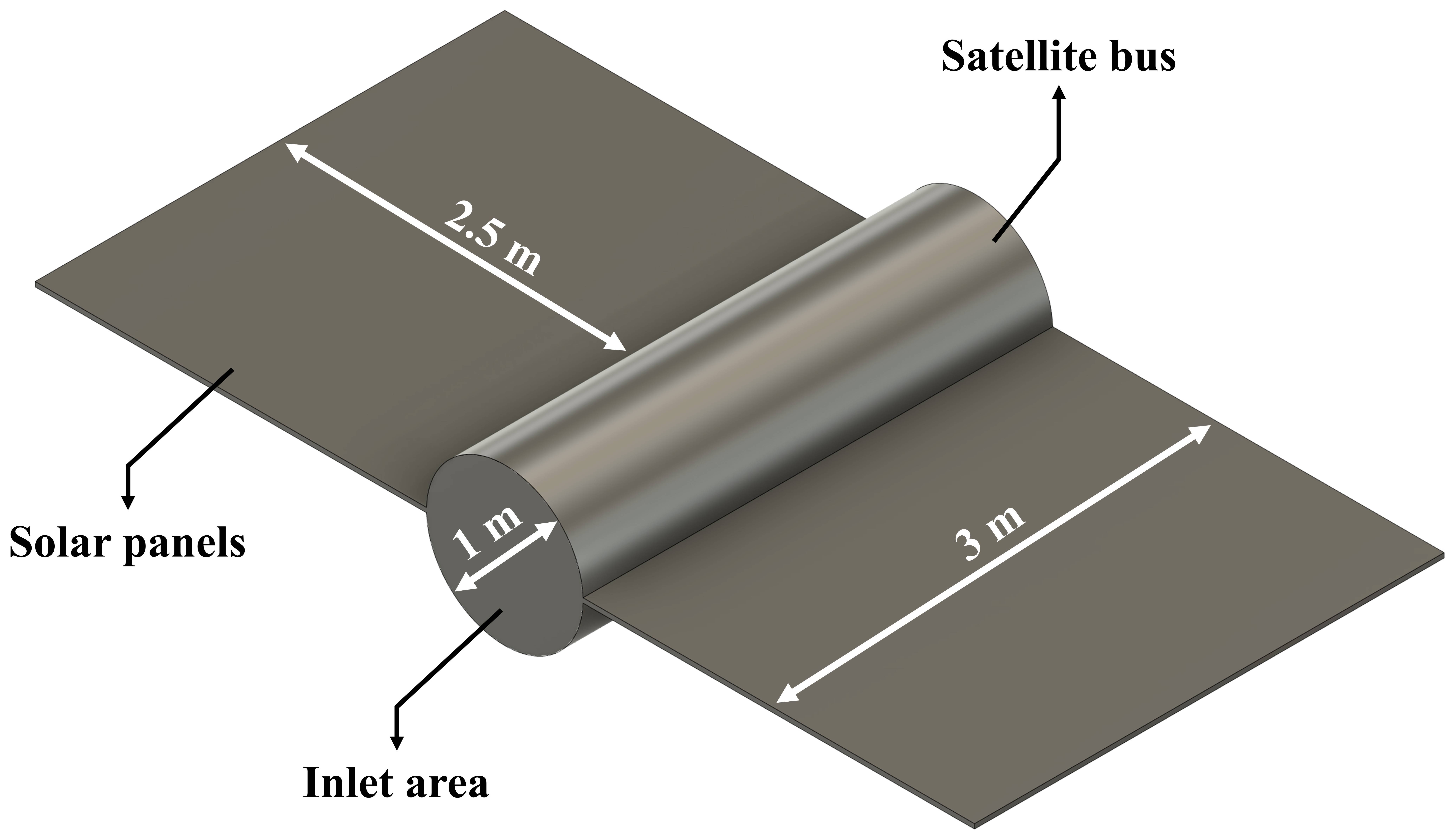}
	\caption{Satellite geometry for a CRAID-integrated ABEP.}
	\label{fig:satellite}
\end{figure}

The satellite geometry in Fig.~\ref{fig:satellite} is developed to investigate the feasibility of drag compensation. The satellite bus has a length of 3 m and a diameter of 1 m, designed to accommodate the dimensions of CPM. A previous study demonstrated that drag calculations using a blocked inlet area can accurately estimate the drag on a satellite incorporating the detailed geometry of the passive intake device, with an error of less than 1\%~\cite{Ko2023}. Hence, the inlet area of the satellite bus is simplified to a flat surface. The solar panels, measuring 3 m in length and 2.5 m in width, are mounted on the exterior surface of the satellite bus. The target orbit is a dawn-dusk sun-synchronous orbit (DDSSO), which minimizes the time the satellite spends in Earth’s shadow. By aligning the solar panels' normal direction with the Sun vector in DDSSO, the effective power-generating area becomes 18 m$^{2}$. With a specific power of 400 W/m$^{2}$, the solar arrays can provide 6 kW for CRAID and the thrusters, leaving a margin of 1.2 kW for payloads and efficiency losses~\cite{Weston2024}. Since $\bar{P}$ calculated in Fig.~\ref{fig:seq_time_power} is less than 6 kW within the altitude range of 150--300 km, the satellite geometry can provide sufficient electric power for CRAID operation. 

\subsubsection{Drag and thrust}
\begin{figure}[htb!] 
	\centering
	\includegraphics[width=9cm]{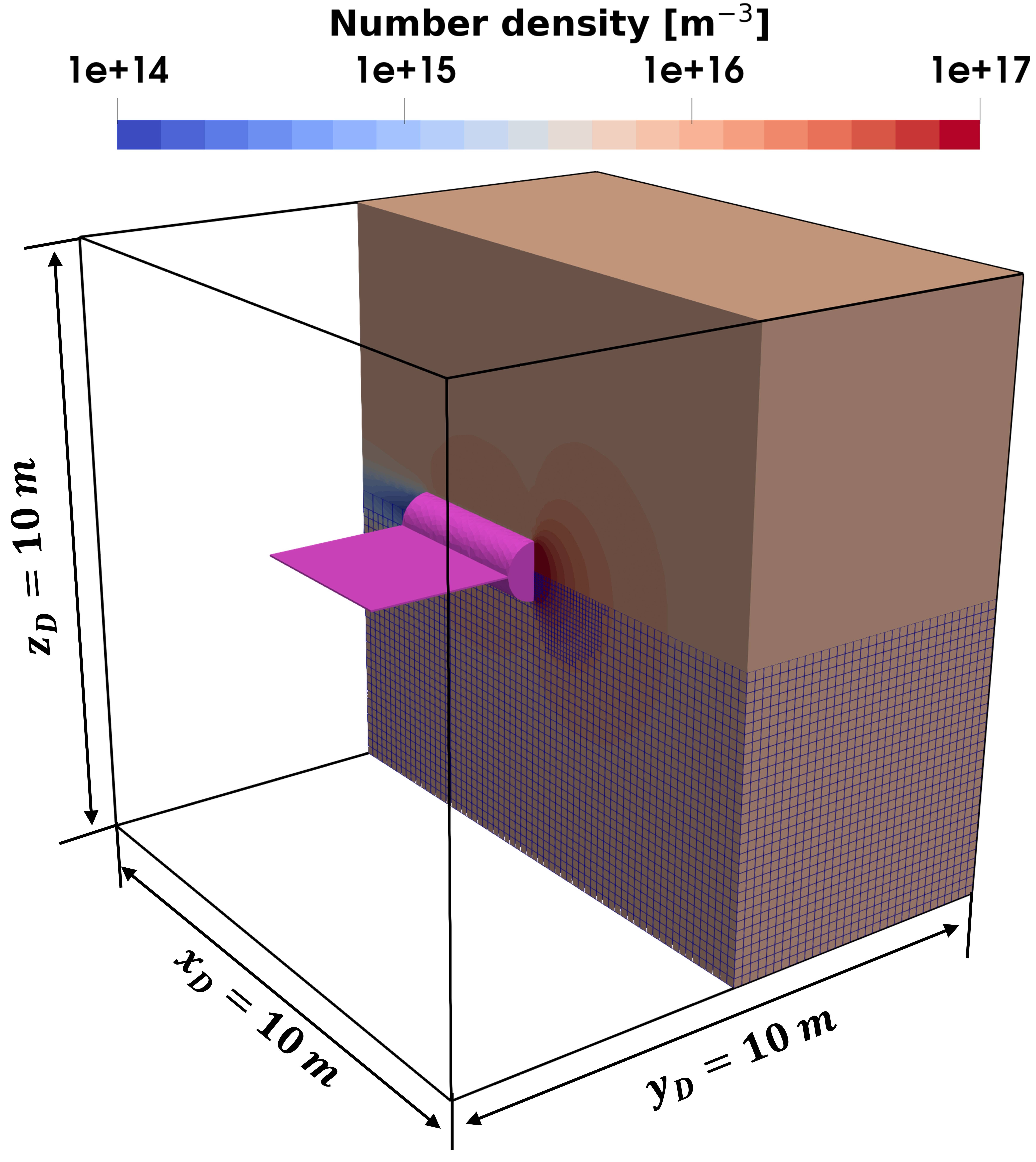}
	\caption{Simulation domain and example number density contours from satellite drag calculation.}
	\label{fig:satellite_domain}
\end{figure}

The drag force $F_D$ acting on the satellite is calculated using the DSMC method. Fig.~\ref{fig:satellite_domain} illustrates the calculation domain and the satellite surface. A diffuse GSI model is applied to the satellite surface, and the freestream is introduced perpendicular to the inlet area. The computational grid is structured using a 3-level octree AMR technique. The time step size is set to $2 \times 10^{-6}$ seconds, and the total drag force acting on the satellite surface is tallied after the flow field reaches a steady state. The calculated drag force at altitudes ranging from 150 km to 300 km is shown as a black symbolic line in Fig.~\ref{fig:alt_force}.\\ 

The thrust $F_T$ is calculated for twin RITs with a grid diameter of 10 cm. The specifications of the RIT are listed in Table~\ref{tab:RIT_spec}. Here, $D_g$ and $D_b$ represent the diameters of the grid and the back wall of the discharge chamber, while $l_c$ denotes its length; $\theta_i$ and $\theta_n$ indicate the grid transparency for ions and neutral particles; $c_f$ represents the Clausing factor; and $V_B$, $N_c$, $R_c$, and $\omega_{RF}$ correspond to the grid voltage, number of coil turns, coil resistance, and RF frequency. The total thruster power is set to 3.5 kW, including both RF power and beam power. The propellant flow rate is assigned as 20 sccm for each thruster, with its composition following the atmospheric conditions at each altitude. Based on the RIT operational conditions, the total thrust from the twin RITs is calculated using the 0D analytical model~\cite{Ko2023}. Here, it is crucial to verify whether the ion beam current density $J$ contributing to thrust exceeds the Child-Langmuir limit $J_{CL}$ to ensure the validity of calculations and the feasibility of thruster operation~\cite{chabert2011physics}:
\begin{equation}
    J_{CL} = \frac{4}{9}\epsilon_{0}\left(\frac{2e}{M_i}\right)^{1/2}\frac{V_{B}^{3/2}}{s^2},
    	\label{eq:CL_limit}
\end{equation}
where $\epsilon_0$ is permittivity of free space, $s$ is the grid spacing, and $\bar{M}$ is the mass of the ions. Assuming $s = 1$ mm, $J_{CL}$ through the open area of the grid is calculated to be 561.3 A/m$^{2}$. In the altitude range of 150–300 km, the $J$ is calculated from the 0D analytical model ranges from 350 to 356 A/m$^{2}$. The current thruster specifications satisfy the Child-Langmuir law for maximum ion beam current density. Based on the calculated $J$, the thrust $F_T$ is estimated as a red symbolic line in Fig.~\ref{fig:alt_force}. The estimated $F_T$ increases from 84.7 mN to 87.1 mN as the chemical composition of the propellant changes with increasing altitude. 

\subsubsection{Flight envelope determination}
\begin{figure}[htb!] 
	\centering
	\subfloat[Forces.]{\includegraphics[width=8.5cm]{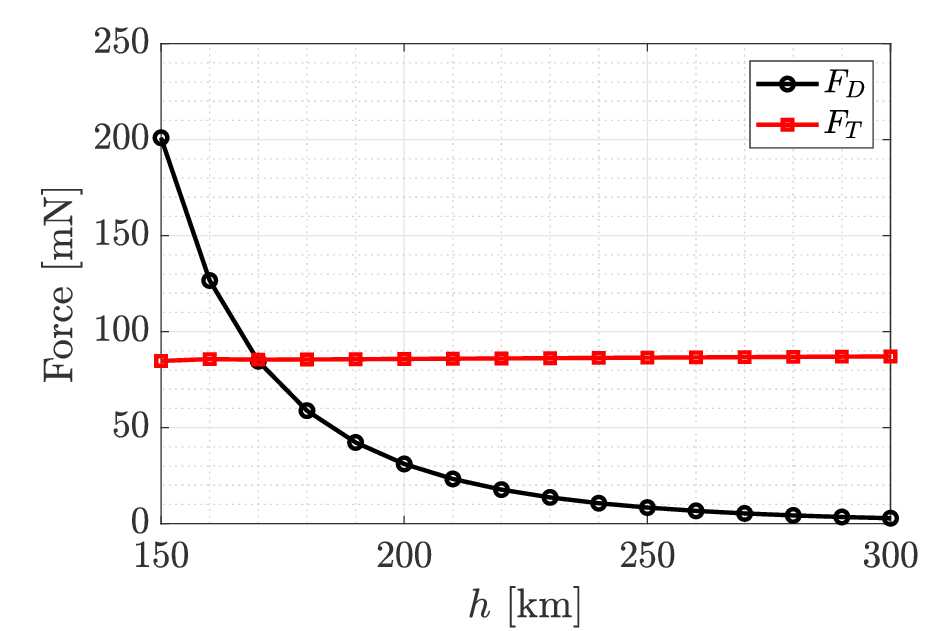}\label{fig:alt_force}}
	\vfill
	\subfloat[Total impulses.]{\includegraphics[width=8.5cm]{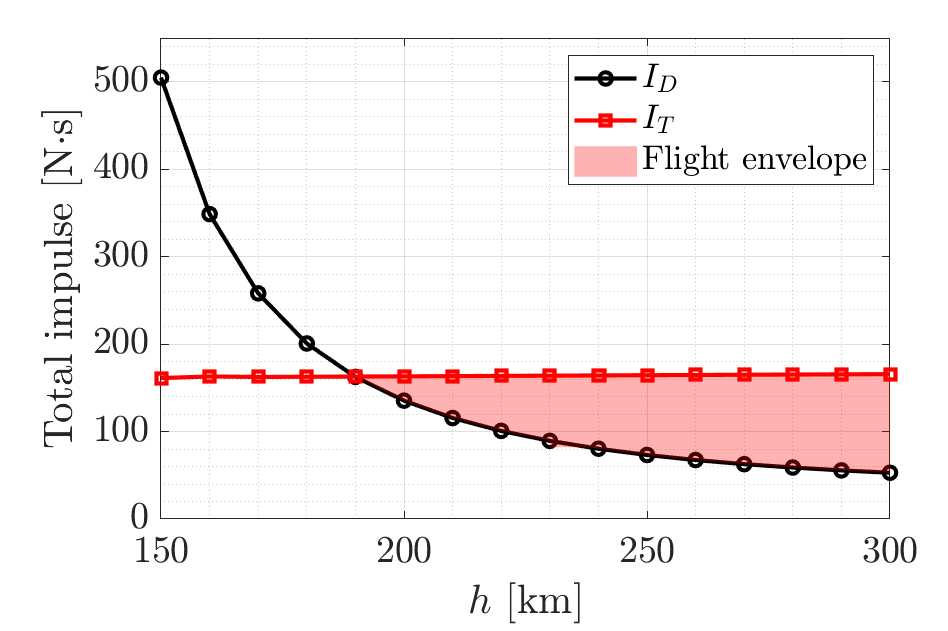}\label{fig:alt_impulse}}
	\caption{Calculated drag and thrust during an operational cycle of CRAID.}
	\label{fig:alt_oper}
\end{figure}

Fig.~\ref{fig:alt_force} shows that thrust exceeds drag at altitudes above 170 km. However, this does not represent the altitude range for complete drag compensation due to the separation of the capturing and propellant utilization processes in CRAID. During $t_{total}^*$ of CRAID operation, thrust is generated only during $t_{regen}^*$, whereas drag acts on the satellite throughout the entire operational cycle. Therefore, the feasibility of drag compensation should be assessed based on the total impulse over the operational cycle. The total drag impulse $I_D$ and thrust impulse $I_T$ for an operational cycle are given by:
\begin{equation}
    I_D = F_D t_{total}^* = F_D (t_{cond}^*+t_{regen}^*+t_{cool}^*),
\end{equation}
\begin{equation}
    I_T = F_T t_{regen}^*.
\end{equation}
Using the forces from Fig.~\ref{fig:alt_force} and sequence duration from Fig.~\ref{fig:seq_time_power}, the total impulse is demonstrated in Fig.~\ref{fig:alt_impulse}. This indicates that the CRAID-RIT integrated system can compensate for the drag acting on the satellite at altitudes above 190 km, which can be defined as the flight envelope. The flight envelope of the CRAID-integrated satellite is not constrained at higher altitudes, unlike conventional passive intake devices~\cite{Ko2023,Tisaev2022}. Passive intake devices may fail to meet the conditions required for thruster operation due to the high degree of atmospheric rarefaction at higher altitudes, which defines the upper boundary of the ABEP flight envelope. In contrast, CRAID decouples the propellant supply process from the capture process, reducing its dependence on atmospheric conditions and enabling an expanded flight envelope for the ABEP system.

\begin{table}
    \centering
    \begin{center}
        \caption{RIT specifications}

        \begin{tabular}{cccccccccc}
        \toprule
        $D_g$ & $D_b$ & $l_c$ & $\theta_i$ & $\theta_n$ & $c_f$ & $V_B$ & $N_c$ & $R_c$ & $\omega_{RF}$ \\ \midrule
        10 [cm] & 5 [cm] & 5 [cm] & 0.85 & 0.62 & 0.6 & 1500 [V] & 10 & 0.2 [$\Omega$]  & 1 [MHz] \\
        \bottomrule 
        \end{tabular}
        \label{tab:RIT_spec}
    \end{center}
\end{table}

\section{Summary and Conclusion}
\label{sec:conclusion}

The intake device plays a crucial role in the operational feasibility of ABEP. CRAID has been proposed as a novel intake device that overcomes the challenges of prevalent passive intake devices by adapting the working principles of a cryopump. This study extends the investigation of CRAID's operability by demonstrating a feasible CPM by assessing system requirements and defining its flight envelope. Numerical methodologies, including the DSMC method with a phase change model and a 0D analytical model for RIT, are employed in this investigation. A significant improvement in intake performance is demonstrated yielding an $\eta_c^{eff}=26.1$\% and a $CR^{eff}=3.0\times10^7$. A comparative analysis of drag and thrust for a model satellite verifies the potential for complete drag compensation. CPM is expected to be operable at altitudes above 190 km. Furthermore, CRAID is capable of delivering a consistent and stable flow rate regardless of environmental conditions, providing an advantage in enhancing the orbital maneuverability of the satellite. However, several challenges remain to be addressed in future work to establish CRAID as a realizable technology. A comprehensive mission analysis and orbit maneuver planning are essential for practical satellite operation, considering that CRAID operates by decoupling atmospheric capture and propulsion processes. Incorporating detailed surface chemistry models for AO-mediated processes could improve the accuracy of CRAID operation simulations, as the current study relies on simplified assumptions for AO recombination~\cite{Huh2025}.


\section{Acknowledgements}
\noindent This work was supported by the Agency For Defense Development by the Korean Government (UG233092TD). This work was supported by the National Supercomputing Center with supercomputing resources including technical support(KSC-2023-CRE-0462).


\end{document}